\documentclass{ws-jnopm}
\usepackage{multirow}
\usepackage{amsmath}
\usepackage{amssymb}
\usepackage{graphics}
\usepackage{graphicx}
\usepackage{array}
\usepackage{color}
\usepackage{chngcntr}
\usepackage[hypertex, colorlinks, colorlinks=false, linkcolor=blue, citecolor=green, filecolor=black, urlcolor=cyan]{hyperref}

\begin{document}

\markboth{Rick Lytel, Shoresh Shafei, Mark Kuzyk}{Optimum topology of \ldots}

\catchline{}{}{}{}{}

\title{OPTIMUM TOPOLOGY OF QUASI-ONE DIMENSIONAL NONLINEAR OPTICAL QUANTUM SYSTEMS.}

\author{RICK LYTEL, SHORESH SHAFEI$^*$, and MARK G. KUZYK}

\address{Department of Physics and Astronomy, Washington State University\\ Pullman, Washington 99164-2814\\
rlytel@wsu.edu}
\address{$^*$Present Address: Department of Chemistry, Duke University, \\ Durham, North Carolina 27708-0346}

\maketitle

\begin{abstract}{We determine the optimum topology of quasi-one dimensional nonlinear optical structures using generalized quantum graph models.  Quantum graphs are relational graphs endowed with a metric and a multiparticle Hamiltonian acting on the edges, and have a long application history in aromatic compounds, mesoscopic and artificial materials, and quantum chaos.  Quantum graphs have recently emerged as models of quasi-one dimensional electron motion for simulating quantum-confined nonlinear optical systems.  This paper derives the nonlinear optical properties of quantum graphs containing the basic star vertex and compares their responses across topological and geometrical classes. We show that such graphs have exactly the right topological properties to generate energy spectra required to achieve large, intrinsic optical nonlinearities.  The graphs have the exquisite geometrical sensitivity required to tune wave function overlap in a way that optimizes the transition moments.  We show that this class of graphs consistently produces intrinsic optical nonlinearities near the fundamental limits.  We discuss the application of the models to the prediction and development of new nonlinear optical structures.}\end{abstract}
%
%

\section{Introduction}\label{sec:intro}

The field of nonlinear optics has spawned decades of fundamental and applied research on the interactions of strong electromagnetic fields with naturally occurring solid,\cite{baugh78.06,lipsc81.01,hugga87.01,cheml80.01} liquid,\cite{giord65.01,giord67.01,ho79.01,chen88.01} liquid crystal,\cite{barni83.01,chen88.01} or gaseous materials,\cite{shelt82.01,kaatz98.01} as well as photonic crystals, mesoscopic solid state \sloppy wires,\cite{guble99.01,foste08.01,tian09.01} and other artificial systems.\cite{cheml85.01,miller1984band,rink89.01,schmi87.01}  Research and commercial developments have driven the search for new materials from which large effects may be extracted with ever-decreasing field strengths and optical intensities.  Ultrafast response is desirable in so many applications that an entire field of research centers on off-resonant effects, whereby photons interact with the material by exciting virtual transitions.  This paper focuses exclusively on a new class of quantum structures for nonlinear optics called quantum graphs, having those energy spectra and wavefunctions required to maximize the off-resonance nonlinear optical response, approaching the fundamental limits allowed by quantum theory.\cite{kuzyk13.01}

\subsection{Ultrafast nonlinear optics}

Nonlinear optical materials are quantum systems with polarizabilities that are nonlinear functions of external electromagnetic fields.  Harmonic generation \cite{maker65.02,bloem68.01,bass69.01}, electro-optics \cite{wayna00.01}, saturable absorption \cite{tutt93.01} , phase conjugation \cite{yariv77.01,yariv78.01}, four-wave mixing \cite{boyd92.01,lytel86.02}, optical bistability \cite{winfu80.01,gibbs84.01}, ultrafast optics \cite{weine11.01}, and waveguide switching \cite{lytel84.01,vanec91.01} are among the many processes in NLO materials \cite{zyss85.01,boyd09.01,kuzyk10.01} of interest in communications, instrumentation, networking, image processing, and many other fields \cite{horna92.01,kuzyk06.06}.

The polarization vector for a general system is a complex function of every allowed transition moment for the material, including electronic, vibrational, and rotational transitions, and their corresponding transition energies and damping factors.  It is often expressed as a power series in the contractions of the nth order susceptibility tensor with n-1 electric field components
\begin{equation}\label{polVector}
P_{i}=\alpha_{ij}\mathcal{E}_{j} + \beta_{ijk}\mathcal{E}_{j}\mathcal{E}_{k} + \gamma_{ijkl}\mathcal{E}_{j}\mathcal{E}_{k}\mathcal{E}_{l} +\ ...
\end{equation}
where $\alpha_{ij}$ is the linear polarizability tensor, $\beta_{ijk}$ is the first hyperpolarizability tensor, $\gamma_{ijkl}$ is the second hyperpolarizability tensor, and the $\mathcal{E}_{j}$ are the external electromagnetic field components to which the material structure couples. For bulk materials, the polarization expansion provides a means to measure the symmetry properties of the susceptibilities and their bulk values.  On the molecular level, the expansion describes the response of a single structure to external optical fields.

Ultrafast nonlinear optical effects are created when the external fields are off-resonance with the energy levels of the system.  The hyperpolarizability tensors become fully symmetric and their magnitudes and rotation properties are entirely determined by the quantum system energy level differences $E_{nm}\equiv E_{n}-E_{m}$ and transition moments $r_{nm}\equiv\left<n|r|m\right>$, where $E_{n}$ are the energies and $n,m$ are state numbers.  They may be calculated using a sum over states.  For the x-diagonal tensor elements, we have\cite{orr71.01}
\begin{equation}\label{betaSOS}
\beta_{xxx}=-3e^3 {\sum_{n,m}}' \frac{x_{0n}\bar{x}_{nm}x_{m0}}{E_{n0}E_{m0}}
\end{equation}
and
\begin{equation}\label{gammaSOS}
\gamma_{xxxx}=4e^4 \left[{\sum_{n,m,l}}' \frac{x_{0n}\bar{x}_{nm}\bar{x}_{ml}x_{l0}}{E_{n0}E_{m0}E_{l0}}-{\sum_{n,m}}' \frac{|x_{0n}|^2|x_{0m}|^2}{E_{n0}^2E_{m0}}\right]
\end{equation}
where $\bar{x}_{nm}\equiv x_{nm}-\delta_{nm}x_{00}$, and the prime on the sums indicates that the ground state $n=0$ is not included.  These will be generalized later to multidimensional space.

The experimentalist desires materials with maximum response in order to minimize the required optical field strength.  Materials are comprised of nonlinear optical structures, or moieties having microscopic hyperpolarizabilities that are calculated from Eq. (\ref{betaSOS}) and Eq. (\ref{gammaSOS}). The size of the moiety may be used to increase its response, but this then limits the number of such moieties that may be incorporated into a material.  It is therefore desirable to create a size-independent metric describing the \emph{intrinsic} nonlinearities.

\subsection{Intrinsic response}

Scale-free, intrinsic hyperpolarizability tensors in the off-resonance regime may be created by normalizing them to their maximum values.  That maxima exist for a given moiety is not immediately clear from the sum over states expressions, but the quantum mechanics of the system provides an additional set of relations among the spectra and transition moments, the Thomas-Reiche-Kuhn sum rules \cite{wang99.01}, viz.,
\begin{equation}\label{TRKsumrule}
S_{nm} = \sum_{p=0}^{\infty} \left[2E_{p0}-(E_{n0}+E_{m0})\right] x_{np}x_{pm}=\frac{\hbar^2 N}{m}\delta_{nm}.
\end{equation}
These constraints set fundamental limits on the hyperpolarizabilities \cite{kuzyk00.01,kuzyk09.01,kuzyk06.03,kuzyk03.01}.  The fundamental limits depend only on the number of electrons, $N$, and the energy gap between the ground and the first excited state, $E_{10}$.  They were first derived by truncating the sum over states Eq. (\ref{betaSOS}) to two levels to calculate a maximum $\beta$, and showing that the addition of a state to the two-level model always reduces the value of $\beta$.  Similar remarks hold for $\gamma$ but with a three level model.  The maxima for the hyperpolarizabilities are then
\begin{equation}\label{sh-betaMax}
\beta_{max} = 3^{1/4} \left(\frac{e\hbar}{m^{1/2}}\right)^3 \frac{N^{3/2}}{E_{10}^{7/2}}
\end{equation}
and
\begin{equation}\label{sh-gammaMax}
\gamma_{max} = 4 \left(\frac{e^4\hbar^4}{m^2}\right) \frac{N^{2}}{E_{10}^{5}} .
\end{equation}

Scale-independent (intrinsic) tensors are created by normalizing each tensor to the fundamental limit.  The fundamental limits are the highest attainable first and second hyperpolarizabilities. Throughout this paper, all tensor components of the hyperpolarizabilities are normalized by these maxima, ie,
\begin{equation}\label{IntrinsicBetaGamma}
\gamma_{ijkl} \rightarrow \frac {\gamma_{ijkl}} {\gamma_{max}} \hspace{2em} \beta_{ijkl} \rightarrow \frac {\beta_{ijkl}} {\beta_{max}} .
\end{equation}
The second hyperpolarizability normalized this way has a largest negative value equal to $-(1/4)$ of the maximum value.  Hyperpolarizabilities normalized this way enable direct comparisons of the intrinsic response without regard to size.

\subsection{Toward an optimum energy spectrum}

It is apparent that structures having the largest nonlinearities will have some set of physical properties making their spectra and transition moments optimum for maximizing the hyperpolarizabilities through the sum over states expressions Eq. (\ref{betaSOS}) and Eq. (\ref{gammaSOS}).  Quantum systems have yet to be found that achieve the maximum allowed values of the response.  In fact, a factor of thirty gap existed between theoretical and experimental values of the first hyperpolarizability prior to 2006 \cite{kuzyk13.01}.  More recent developments have led to molecules with record hyperpolarizabilities \cite{zhou06.01,zhou07.02,perez07.01,perez09.01}, still well short of the fundamental limits.  Apparently the spectra and transition moments $(E_{n0},x_{nm})$ for all of these moieties are far from optimum.

\begin{figure}\centering
\includegraphics[width=2.5in]{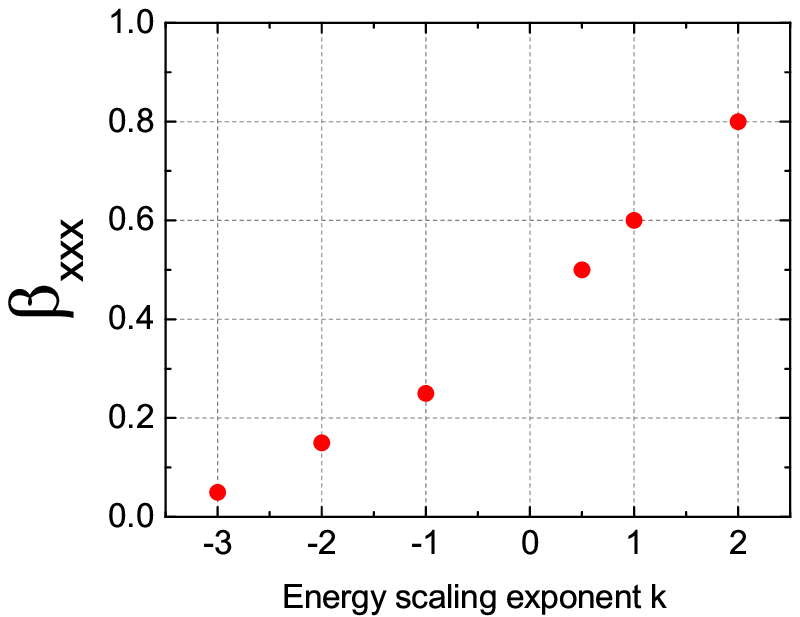}\includegraphics[width=2.5in]{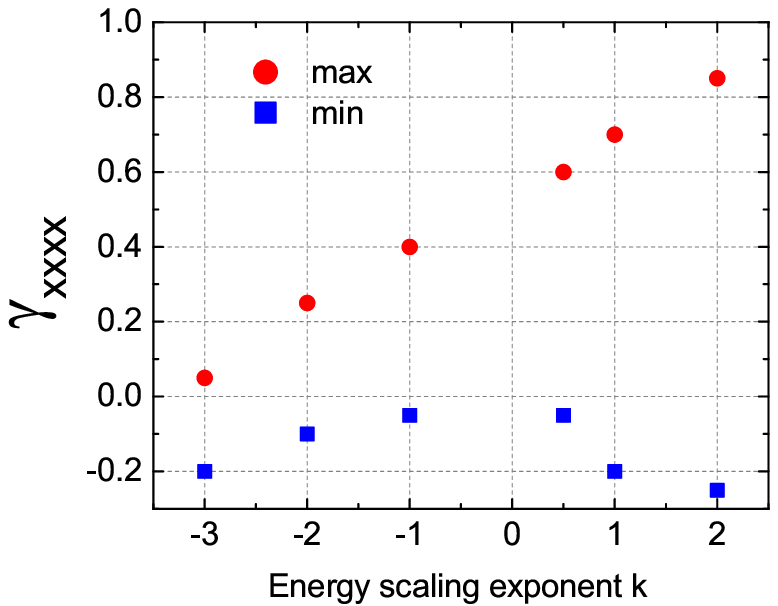}\\
\caption{Scaling of the maximum intrinsic first hyperpolarizability (left) and the maximum and minimum intrinsic second hyperpolarizability with the scaling exponent k, where the energy spectrum scales with state number n as $E_{n}\sim n^{k}$.}
\label{fig:MClimits}
\end{figure}

This observation is borne out through theoretical explorations of the behavior required of states and spectra for a general quantum system using Monte Carlo methods, constrained by the full TRK sum rules \cite{bello08.01,wang99.01,kuzyk06.01}. Monte Carlo studies of large numbers of constrained sets of spectra and moments reveal that the optimum spectrum for a system scales quadratically or faster with state number (which we call a superscaling spectrum) \cite{shafe11.01}.  Fig. \ref{fig:MClimits} shows how the intrinsic hyperpolarizabilities scale with the state scaling exponent k, defined through the scaling of the energy $E_{n}$ with state number n as $E_{n}\sim n^{k}$.  Topological properties of a system determine the spectra, while geometrical properties determine projections of the transition moments onto a specific external axis.  An optimum topology is one which produces a superscaling spectrum.

The Monte Carlo results are precise, but a heuristic from the theory of fundamental limits will be shown to corroborate them.  The three-level Ansatz (TLA), which posits that only three states contribute for a system with a nonlinearity close to its limit, is consistent with all observations and analysis to date, and is a suitable heuristic to consider:
\begin{eqnarray}\label{betaInt3Lmodel}
\beta_{xxx}\rightarrow\beta_{xxx}^{3L} &=& 3^{3/4}\left(\frac{m}{\hbar^2}\right)^{3/2}E_{10}^{7/2} \left[\frac{|x_{01}|^{2}\bar{x}_{11}}{E_{10}^{2}}\right. \\
&+& \left.\frac{|x_{02}|^{2}\bar{x}_{22}}{E_{20}^{2}} +\left(\frac{x_{01}x_{20}x_{12}}{E_{10}E_{20}}+c.c.\right)\right]\nonumber
\end{eqnarray}
(with an equivalent expression for $\gamma_{xxxx}$, but requiring four states).  Note that the TLA does not imply that only three states are required to obey the sum rules. Contributions of low lying states may dominate the hyperpolarizabilities, but contributions from large numbers of states to the sum rules are usually required.  All known models with a given topology employing a potential energy satisfy the TLA when their geometries are tuned to yield the maximum hyperpolarizabilities for their topology.\cite{lytel13.01}

Suppose we now simplify the three level model by replacing the products of moments in Eq. (\ref{betaInt3Lmodel}) with expressions computed from the lowest four sum rule constraints Eq, (\ref{TRKsumrule}) truncated to only three levels.  Doing so leads to the \emph{extreme} three level (3L) model \cite{kuzyk00.01},  a two parameter expression for $\beta$ in this limit:
\begin{eqnarray}\label{betaIntfG}
\beta^{3L}_{xxx}&\rightarrow& \beta^{3L}_{xxx}(ext)=f(E)G(X),\nonumber \\
f(E)&=&\left[(1-E)^{3/2} \left( E^2 + \frac {3} {2} E + 1 \right)\right], \\
G(X)&=&\left[\sqrt[4]{3} X \sqrt{\frac {3} {2} \left( 1 - X^4\right)}\right]\nonumber
\end{eqnarray}
with
\begin{equation}\label{Xmax}
E\equiv E_{10}/E_{20},\ E_{n0}=E_{n}-E_{0}
\end{equation}
and
\begin{equation}\label{Xmax}
X\equiv x_{01}/x_{01}^{max},\ x_{01}^{max} = \left(\frac{\hbar^2 N}{2 m E_{10}}\right)^{1/2}.
\end{equation}

\begin{figure}\centering
\includegraphics[width=4in]{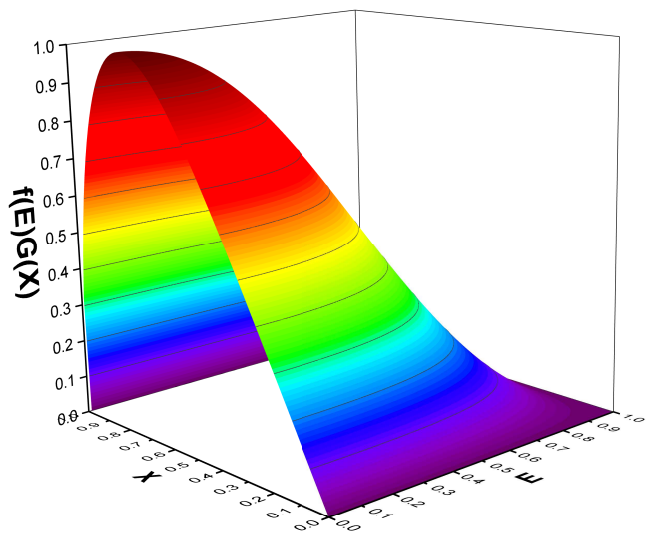}
\caption{Scaling of the extreme three level (intrinsic) first hyperpolarizability with the three level model parameters X and E.  The maximum of unity occurs when $E=0$ and $X=0.79$.}\label{fig:XEfGsurfaceplot}
\end{figure}

Figure \ref{fig:XEfGsurfaceplot} shows the dependence of the extreme three level $\beta^{3L}_{xxx}(ext)=f(E)G(X)$ on the two parameters X and E.  As noted above, derivation of these limits requires truncation of the four lowest sum rules to three states, and if this truncation is to be exact, a physical system would have to have moments and spectra satisfying ancillary constraint conditions, viz.,
\begin{equation}\label{TRKsumruleTrunc3}
\sum_{p=3}^{\infty} \left[2E_{p0}-(E_{n0}+E_{m0})\right] x_{np}x_{pm}=0
\end{equation}
for $(n,m)=(0,0), (0,1), (0,2)$, and $(1,2)$, a seemingly improbable event.  Also, higher order sum rules, eg, $S_{22}$ would necessarily require more than three states to converge.

The extreme three level model is the asymptotic limit of the TLA as the three level sum rules become exact, ie, as $\beta\rightarrow  1$.  In this limit, the scaling parameters take the values $(E,X)\rightarrow (0,0.79)$.  As $E\rightarrow 1$, the model predicts that $\beta \rightarrow 0$. In fact, if $E_{n}\sim 1/n^2$, then the ratio $E\sim 27/32$, suggesting that molecular structures with Coulomb-like potentials should have poor first hyperpolarizabilities, as indeed they all do. Though the model is only approximate when E moves away from zero, it suggests that energies scaling inversely with an inverse power of the state number should have very small $\beta$, while those scaling with some positive power of the state number should have larger $\beta$, thus corroborating the Monte Carlo results.

It has been established that maximizing the hyperpolarizabilities is of little use in determining the potential in any model \cite{ather12.01,burke13.01}, an important conclusion and one that suggests a variety of potentials comprised of continuous, piecewise continuous and even discontinuous functions might generate a large response, so long as the spectrum is superscaling.  Note that superscaling spectra need not be regular, ie, the energies may fall between fixed boundaries that superscale with state number, so-called root boundaries.

We also note that all nonrelativistic single and many particle Hamiltonians studied to date have spectra and states that generate nonlinearities that are at most $70\%$ of the maximum for $\beta$.  For the second hyperpolarizability, the apparent limits are $60\%$ of the fundamental limits.  These appear to be hard limits.  It is not known whether systems exist that can achieve the actual fundamental limits. \cite{kuzyk13.01}.  But even these so-called potential limits are desirable to achieve, as they are much larger than anything known today.

In summary, we require systems with superscaling spectra in order to have any chance of achieving a significant fraction of the theoretical maxima of the nonlinear optical response.  We also require systems with geometries that can provide transition moments that maximize the response while simultaneously being constrained by the TRK sum rules.  We now turn to quantum graphs, which are dynamical models of systems that exhibit the desired spectral scaling behavior with a rich range of tunable geometries for optimizing the transition moments as well.

\subsection{Quantum graph models}

With these observations, a search was begun to discover model systems with superscaling spectra that could be reasonably expected to reflect the physics of an optimum nonlinear optical structure.  We settled on quantum graphs as a near-ideal model system.  Quantum graphs are defined by starting with a quasi-one dimensional relational graph \cite{bollo98.01,newma03.01}, giving it a metric, allowing an operator to act on the edges, and providing boundary conditions ensuring the system satisfies completeness, closure, and unitarity. A quantum graph (QG) defined this way is a general confinement model for quasi-one dimensional electron dynamics.

Quantum graphs are idealized models of various physical and chemical systems, devised to gain insight into problems that were otherwise analytically intractable.  They were introduced by Pauling in his study of the diamagnetic anisotropy of aromatic compounds \cite{pauli36.01}.  In this model, electrons move from one carbon atom to another along the bonds under the influence of external fields.  Generalizations were created by Kuhn \cite{kuhn48.01}, and by Ruedenberg and Scherr \cite{ruede53.01} as a free electron model for conjugated systems.  This led to a series of papers of numerical calculations \cite{scher53.01}, demonstration models \cite{platt53.01}, and a rigorous mathematical formulation of the model.

Quantum graphs have subsequently been investigated for applications to mesoscopic systems \cite{kowal90.01}, optical waveguides \cite{flesi87.01}, quantum wires \cite{ivche98.01,sanch98.01}, excitations in fractals \cite{avish92.01}, and fullerines, graphene, and carbon nanotubes \cite{amovi04.01,leys04.01,kuchm07.01}. Kuchment\cite{kuchm02.01} provides a concise survey of the application of quantum graphs to thin structures, including photonic crystals.  In recent years, quantum graphs were shown to be exactly solvable models of quantum chaos \cite{kotto97.01,kotto99.02,kotto00.01,blume01.04}.  Comprehensive reviews of this enormous field are available in the literature \cite{exner08.01,berko06.01,berko13.01}, and the mathematical rigor for these models is quite impressive.  A recent, thorough review of applications of quantum graphs may be found in Chapter 7 of Berkolaiko and Kuchment\cite{berko13.01}.

For all of these models, the mathematics of reduction of a complex physical problem to that of particle dynamics governed by a differential operator on the edge of a graph is itself intractable, so heuristic arguments are often employed to justify the use of the models.  Also, models with finite transverse dimensions and leaky modes have been studied to improve the realism of the model.  As with any such endeavor, a great deal about the original system may be learned by studying a simplified model, including ways to improve the model and attempt to generate experimental confirmation.  Microwave networks have recently been successfully used to experimentally simulate quantum graphs \cite{hul04.01}.

Most recently, quantum graphs have emerged as models of tightly confined nanowire systems for nonlinear optics \cite{shafe12.01,lytel13.01,lytel13.04} because of their superscaling energy spectra, and their large number of topological and geometrical configurations.  Quantum graph models for nonlinear optics were initially studied to explore the effects of extreme confinement in quantum wires \cite{shafe11.02,shafe12.01}.  In these models, motion in the dimension transverse to the wire is confined by a limiting procedure, and the electrons are confined to dynamics along the wire.  In the limit of vanishing transverse width, only the longitudinal motion contributes to the hyperpolarizabilities.  However, the TRK sum rules contain contributions from both the longitudinal and transverse dimensions.  The existence of new nonlinear optical physics on quantum confined wires is the genesis of the quantum graph explorations, as such graphs have a rich range of topologies and geometries with which to explore optimization of the response.

The generalized QG model of an $N$ electron structure constrains dynamics to the edges of the metric graph.  Dynamics are governed by a self-adjoint, multiparticle Hamiltonian operating on the edges and possessing a complete set of eigenstates and eigenvalues.  The general Hamiltonian contains momentum, position, and spin operators, as well as functions of each describing particle-particle interactions, coupling to external fields, and other interactions.  Transitions between states determine the nonlinear optical response of the graph to an external optical field.  The canonical commutation relations guarantee that the TRK sum rules hold for the transition moments, providing constraints and relations among the various allowed transitions in the system \cite{kuzyk00.01,kuzyk06.03}.

The one-electron version of the generalized quantum graph model (hereafter referred to as the \emph{elementary QG}) is exactly solvable.  Quantum graphs with zero potential energy (bare edges) and nonzero potentials (dressed edges) have recently been applied to calculate the off-resonance first ($\beta_{ijk}$) and second ($\gamma_{ijkl}$) hyperpolarizability tensors (normalized to their maximum values) of elementary graphical structures, such as wires, closed loops, and star vertices \cite{lytel12.01,shafe12.01} and to investigate the relationship between the topology and geometry of a graph and its nonlinear optical response \cite{lytel13.01} through its hyperpolarizability tensors.  The results showed that the elementary QG model of a 3-edge star graph generated a first hyperpolarizability over half the fundamental limit and a second hyperpolarizability whose range was between $20-40\%$ of the fundamental limit.  These results suggest that graphs comprised of combinations of stars could be suitable models of moieties with the superscaling spectra required to approach the fundamental limits, the subject of this paper.

It is desirable to use the exactly solvable elementary QG models for exploring the topologies yielding the largest hyperpolarizabilities, mainly because they are simpler and tractable compared to a generalized QG model.  That this may be done is seen as follows.  The generalized QG model may be expected to reflect dynamics of a multi-electron system, e.g., a solid nanowire with a band structure and a superscaling spectrum and set of transition moments satisfying the TRK sum rules.  Since the topology of the generalized QG determines the scaling of the spectrum, we expect the spectrum of the generalized and elementary QG models to be globally similar, though differing in details of the excited-state spectrum arising from electron interactions.  The transition moments of the generalized QG are sums of the moments for each electron, and it should be expected that the geometry of the generalized QG can be manipulated to more or less match the moments of some geometry of the single electron elementary QG.  This implies that the spherical tensor components of both the generalized and elementary QG should exhibit similar global behavior with the shape of the nanowire.  Computations of the hyperpolarizability of multielectron systems show that their maxima are unchanged from those of the single electron systems when the number of electrons is accounted for.\cite{watki11.01}  Thus, the use of the elementary QG model to investigate the optimum topology for a nonlinear optical nanostructure is a physically meaningful step.

Section \ref{sec:QGreview} reviews the general methods for calculating the required spectra and transition moments for quasi-one dimensional graphs, and the first and second hyperpolarizability tensors in the quasi-1D limit, i.e., when the two-dimensional calculation is reduced by taking the transverse dimension along the edges of the graph to zero while confining the electrons with an infinite potential.  The concept of a graphical motif is introduced and discussed.  Section \ref{sec:topOpt} presents the hyperpolarizabilities for a variety of star-based topologies and shows how this class of graphs can approach over $60\%$ of the fundamental limit.  It is also shown how a change in topology can cause a dramatic change in the response by altering the states and spectrum contributing to the hyperpolarizabilities. Section \ref{sec:outlook} summarizes the application of the elementary QG model to elementary and composite graphs and points to the next direction.  Several Appendices contain the detailed methods for calculating quantum graph states and spectra, show how to solve graphs when the spectra are degenerate, and show how to scale the star motif to $N\geq 4$ edges.

Two new and fundamental results emerge from this work that can aide the molecular designer of nanowire and quantum-confined systems for nonlinear optics.  The first is that the optimum topology for quantum confined systems are those containing star vertices, and that wires and loops are simply not capable of achieving a large response.  The second is that the star topologies generally have the largest intrinsic responses achievable to date and might be realized in quasi-one dimensional nanostructures having superscaling spectra and a broad range of possible transition moments.

\section{Nonlinear optics in the elementary QG model}\label{sec:QGreview}

\subsection{A quasi-1D dynamical system}
The dynamics of an electron on a quantum graph are described by a self-adjoint Hamiltonian operating on the edges of the graph, with complex amplitude and probability conservation (hereafter referred to as flux conservation throughout the paper) at all internal vertices and fixed, infinite potentials at the termination vertices (where the amplitude vanishes).  The physics of the eigenstates and their spectra have been previously described, along with a suitable lexicography for describing the \emph{union} operation for creating eigenstates from the edge functions that solve the equations of motion for the Hamiltonian \cite{shafe12.01}.

\begin{figure}\centering
\includegraphics[width=3.4in]{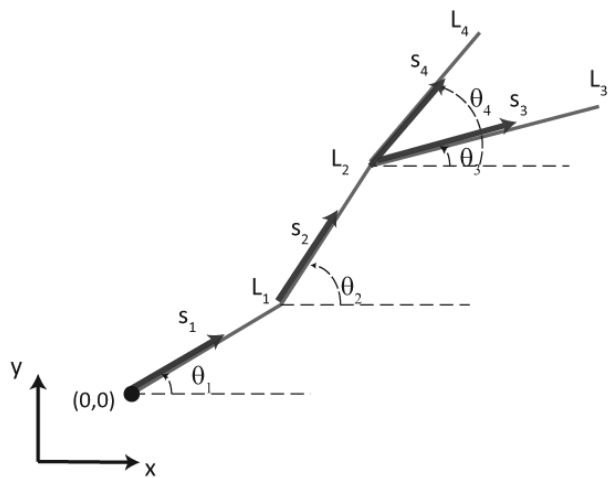}
\caption{A four-edge quantum graph.  Each edge has its own longitudinal coordinate $s_{i}$ ranging from zero to $L_{i}$.  The projection $x(s_{i})$ of an edge onto the $x-axis$ is measured from the origin of the coordinate system attached to the graph (and not to the beginning of the edge itself).  For example, $x(s_1)=s_1\cos\theta_1$ while $x(s_2)=L_1\cos\theta_1 + s_2\cos\theta_2$ and so on.}\label{fig:graphNEW}
\end{figure}

The graph is specified by the location of its vertices and the edges connecting the vertices.  Fig \ref{fig:graphNEW} details the notation and configuration of a graph.  A set of vertices with arbitrary locations in the 2D plane but fixed connections specifies a topological class of graphs.  For a fixed topology, the variation of vertex locations specifies various geometries for the graph.  Since motion is confined to the graph edges and is continuous at each vertex, the energy spectrum depends only on the edge lengths and the boundary conditions, ie, the topology.  Spectra are quasi-quadratic in state number, ie, superscaling.  This is the first requirement for a large response from a nonlinear optical structure.

The edge lengths and angular positions determine the projections of electron motion onto a fixed, external reference axis.  The projections summed over all edges yield the transition moments required to compute the tensor elements of the hyperpolarizability tensors.  Regardless of how the axes used to define the vertices are chosen, the various tensor components may be used to assemble any component in a different frame by using the rotation properties of the tensors.

The study of the nonlinear optical properties of a specific graph topology requires solving the graph for its eigenstates and spectra as functions of its edge lengths and using them to compute a set of transition moments for the graph from which the hyperpolarizability tensors may be computed.

\subsection{States and spectra}\label{statesAndSpectra}

We address so-called bare graphs in this paper.  These are graphs with zero potential energy everywhere except at their external vertices, where the potential is infinite.  A set of dressed graphs having potentials on the edges has recently been analyzed for nonlinear optics.\cite{lytel13.04}.

A general quantum graph is solved for its spectra and states once the Hamiltonian has been specified.  Then, the Hamiltonian is used to compute a set of edge functions, $\phi_n^{i}(s_{i})$ that are solutions to the Schr\"{o}dinger equation with eigenvalues $E_{n}$, which will be the same on all edges and identical to those for the entire graph.  Next, the edge functions are used to construct eigenstates of the Hamiltonian for the graph through union process that reflects a direct sum Hilbert space over the edges \cite{shafe12.01}:
\begin{equation}\label{eigenFunction}
\psi_n(s)= \cup_{i=1}^{E} \phi_n^{i}(s_{i})
\end{equation}

The edge functions for an edge connecting a vertex with amplitude $A_{n}$ to vertex with amplitude $B_{n}$ may be written in a canonical form that automatically matches (nonzero) amplitudes at each internal vertex:
\begin{equation}\label{edgeFunctionAB}
\phi_n^{i}(s_i)= \frac {A_{n}^{(i)} \sin k_{n}(a_i-s_i)+B_{n}^{(i)}\sin k_{n}s_i}{\sin k_{n}a_i}
\end{equation}

Vertices with zero amplitude occur when the edges in the graph are rationally-related and are discussed in Section \ref{sec:motif}.  For the rest of this paper, we assume the edges are irrationally-related, so that the denominator in Eq. (\ref{edgeFunctionAB}) never vanishes.  Terminal vertices with zero amplitude will take the form of Eq. (\ref{edgeFunctionAB}) with one of $A_{n}$ or $B_{n}$ equal to zero.

The union is defined such that the eigenstate is continuous at every vertex of the graph, while the probability current is conserved at each vertex and thus throughout the graph.  These two boundary conditions guarantee that the eigenstates are complete.  They also generate the relationships among the edge amplitudes required to compute the spectrum of the graph.

There are always a sufficient number of equations among the coefficients such that a characteristic equation for the eigenvalues may be extracted.  To see this, let the graph have E edges, $V_{E}$ external vertices, and $V_{I}$ internal vertices, each internal vertex $V_{i}$ having a degree $d_{i}$ that counts the number of edges connected to vertex $V_{i}$.  The total number of vertices is $V=V_{E}+V_{I}$. The number of boundary conditions needs to be identical to the number of unknown coefficients, which is $2E$ (two for each edge, according to Eq. (\ref{edgeFunctionAB}).  There are exactly $V_{E}$ amplitude boundary conditions from the external vertices.  There are also exactly $V_{I}$ flux boundary conditions from the internal vertices.  So far we have exactly $V$ boundary conditions.  At each vertex having degree $d_{i}$, there are exactly $d_{i}^{k}-1$ amplitude boundary conditions, and summing this over all internal vertices yields $\sum_{i}(d_{i}-1)$ total amplitude boundary conditions.  The first sum is simply the number of edge endpoints connected by internal vertices, and is clearly equal to $2E-V_{E}$, while the second sum is the number of internal vertices, $V_{i}$.  This yields exactly $2E-V_{E}-V_{I}=2E-V$ amplitude boundary conditions at internal vertices. The total number of boundary conditions is obviously $2E$, as required.

Solutions of the 2E coupled amplitude equations resulting from the boundary conditions exist only if the determinant of the matrix of coefficients vanishes.  This condition produces the secular or characteristic equation for the graph and determines the eigenvalues $k_{n}$ and the exact energy spectrum $E_{n}=\hbar^{2}k_{n}^{2}/2m$.  Since the boundary conditions in the elementary QG model are independent of the angles the edges make with respect to one another, the secular equation is independent of angles and depends only on dimensionless parameters $k_{n}a_{i}$.  For a given configuration of vertices, the distance between them and the rules by which they are connected, i.e., the topology of the graph, determines the energy spectrum.  Topologically different graphs with identical geometries have different energy spectra.  In this way, the graph topology has a large impact on the nonlinear optical response.

Except for bent wires and closed loops, the secular equation of a graph is generally a transcendental equation.  Accurate solutions are easily found numerically.  From these, the internal amplitudes $A_{n}^{(i)}$ and $B_{n}^{(i)}$ may be calculated relative to the same normalization constant.  Normalizing the eigenfunction produces the states required to compute the transition moments.

It should be noted that the transition moments are sums (not unions) over edges of the following form:
\begin{equation}\label{xNM}
x_{nm}=\sum_{i=1}^{E}\int_{0}^{a_{i}}\phi_{n}^{*i}(s_{i})\phi_{m}^{i}(s_{i})\ x(s_{i})ds_{i}
\end{equation}
where $\phi_{m}^{i}(s_{i})$ is the normalized wave function on the $i^{th}$ edge that obeys the boundary conditions for the graph and $x(s_{i})$ is the x-component of $s_{i}$, measured from the origin of the graph (and not of the edge), and is a function of the prior edge lengths and angles.  With edge wave functions of the form in equation (\ref{edgeFunctionAB}), the computation of the transition moments requires integrals of products of sines and cosines with either $s$ or $1$, all of which are calculable in closed form \cite{shafe12.01,lytel13.01,lytel12.01}.

We note that the transverse wavefunction is not calculated and not essential in this model.  The transverse states do not contribute to the hyperpolarizabilities \cite{shafe11.02}.  But there are residual effects from the transverse state in the sum rules, as has been previously discussed \cite{shafe11.02,shafe12.01}.

The process for calculating the spectra and transition moments of elementary QG's can be summarized as follows: (1)  select a particular graph topology, specifying the number of vertices and the connecting edges, (2)  generate a random set of vertices, and calculate the lengths of the edges and the angles each makes with the $x$-axis of the graph's coordinate system, (3) solve the Schr\"{o}dinger equation on each edge of the graph, and (4) match boundary conditions at the vertices and terminal points.  This results in a set of equations for the amplitudes of the wavefunctions on each edge.  The solvability of this set requires that the determinant of the amplitude coefficients vanishes, leading to a secular equation for the eigenvalues.  The transition moments $x_{nm}$ and energies $E_{n}=\hbar^{2}k_{n}^{2}/2m$ may be used to compute the first and second hyperpolarizabilities of any graph specified by a set of vertices, as described next.

\subsection{Hyperpolarizability tensors}\label{tensorCalcs}

For the quasi-1D problem the tensors are indexed in the (x,y) directions.  The full tensor expressions are given as a sum over states.  We choose to normalize energies and transition moments to provide a concise, dimensionless expression for both hyperpolarizabilities.  The first intrinsic hyperpolarizability tensor for 2D graphs may then be written as
\begin{eqnarray}\label{betaInt}
\beta_{ijk} &\equiv& \frac{\beta}{\beta_{max}} = \left(\frac{3}{4}\right)^{3/4} {\sum_{n,m}}' \frac{\xi_{0n}^{i}\bar{\xi}_{nm}^{j}\xi_{m0}^{k}}{e_n e_m} ,
\end{eqnarray}
where $\xi_{nm}^{i}$ and $e_n$ are normalized transition moments and energies, defined by
\begin{equation}\label{xNMnorm}
\xi_{nm}^{i} = \frac{r_{nm}^{i}}{r_{01}^{max}}, \qquad e_{n} = \frac{E_{n0}}{E_{10}},
\end{equation}
with $r^{(i=1)}=x$ and $r^{(i=2)}=y$, $E_{nm}=E_n-E_m$, and where
\begin{equation}\label{Xmax}
r_{01}^{max} = \left(\frac{\hbar^2}{2 m E_{10}}\right)^{1/2}.
\end{equation}
$r_{01}^{max}$ represents the largest possible transition moment value of $r_{01}$ \cite{kuzyk00.01}.  According to equation (\ref{xNMnorm}), $e_0 = 0$ and $e_1 = 1$. $\beta_{ijk}$ is scale-invariant and can be used to compare molecules of different shapes and sizes.
Similarly, the second intrinsic hyperpolarizability is given by
\begin{eqnarray}\label{gammaInt}
\gamma_{ijkl} &=& \frac{1}{4} \left({\sum_{n,m,l}}' \frac{\xi_{0n}^{i}\bar{\xi}_{nm}^{j}\bar{\xi}_{ml}^{k}\xi_{l0}^{l}}{e_n e_m e_l} - {\sum_{n,m}}' \frac{\xi_{0n}^{i}\xi_{n0}^{j}\xi_{0m}^{j}\xi_{m0}^{k}}{e_n^2 e_m}\right) . \nonumber \\
\end{eqnarray}

We already know that quantum graphs exhibit superscaling spectra and that some topologies are expected to have shapes which yield large nonlinearities.  Which geometries of a given topology yield a larger response? And which topologies show the most promise for enabling a specific geometry to have one of the larger possible responses?  This knowledge is obtained by specifying a fixed topology, such as a star, loop, or wire, and calculating the response for a large number of possible geometries in order to discover the \emph{best} shape.  By best, the experimentalist usually means the one with the largest value of the hyperpolarizability in a lab frame whose $x$-axis is known and usually used to reference the optical field polarizations interacting with the material.

The specification of a graph through its vertices, the calculation of its states and spectra, and the sampling of large numbers of geometries to create ensembles of transition moments, energies, and hyperpolarizabilities, is a Monte Carlo computation.  The results of such a calculation are a set of tensors for a topological class of graphs whose variability is solely determined by the geometrical properties of the graphs.

The graphs are generated by randomly picking vertices and connecting them to generate the desired topology.  The transition moments and hyperpolarizability tensors are computed in the reference frame defined by the coordinate system used to specify the vertices.  The four nonzero tensor components for $\beta_{ijk}$ and the five nonzero tensor components for $\gamma_{ijkl}$ represent the tensors of the graph in the x-y frame.  If x is the laboratory x-axis, then the initial graph is likely not in the correct orientation to yield the largest x-diagonal components desired by the experimentalist.  But the graph is easily rotated through an angle into the orientation yielding the largest diagonal components by using the rotation properties of the tensors:

\begin{eqnarray}\label{BetaCartesian}
\beta_{xxx}(\phi) &=& \beta_{xxx}\cos^3\phi \nonumber + 3\beta_{xxy}\cos^2\phi \sin\phi \nonumber \\ &+&  3\beta_{xyy}\cos\phi \sin^2\phi + \beta_{yyy}\sin^3\phi,
\end{eqnarray}
and
\begin{eqnarray}\label{GammaCartesian}
\gamma_{xxxx}(\theta) &=& \gamma_{xxxx}\cos^4\theta+4\gamma_{xxxy}\cos^3\theta\sin\theta\nonumber \\
&+& 6\gamma_{xxyy}\cos^2\theta \sin^2\theta+4\gamma_{xyyy}\cos\theta \sin^3\theta\nonumber \\
&+& \gamma_{yyyy}\sin^4\theta ,
\end{eqnarray}
where the values of $\phi$ and $\theta$ that maximize the left-hand side of either equation usually differ, and the tensor components on the right-hand side of either equation are referenced to zero rotation angle, ie, the original position of the graph.  By definition, $\beta_{xxx}(\phi)$ ($\gamma_{xxxx}(\theta)$) is at an extreme value when the graph is rotated through $\phi$ ($\theta$).  Once the graph is solved and the tensor components are known in its frame, $\phi$ ($\theta$) is easily found by maximizing equation (\ref{BetaCartesian}) for $\beta_{xxx}(\phi)$ (equation (\ref{GammaCartesian}) for $\gamma_{xxxx}(\theta)$).

The tensor norms are the full contraction of the tensors with themselves and are invariant under any transformation by the rotation group, and provide immediate insight into the limiting responses of the graphs.  They are given by
\begin{equation}\label{BetaNorm}
|\beta| = \left( \beta_{xxx}^2+3\beta_{xxy}^2+ 3\beta_{xyy}^2+\beta_{yyy}^2\right)^{1/2}
\end{equation}
and
\begin{equation}\label{GammaNorm}
|\gamma| = \left( \gamma_{xxxx}^2+4\gamma_{xxxy}^2+ 6\gamma_{xxyy}^2+4\gamma_{xyyy}+\gamma_{yyyy}^2\right)^{1/2} .
\end{equation}
These are the magnitudes of the graph's hyperpolarizabilities and are both scale and orientation-independent. The use of tensors to extract the nonlinear optical response as a function of geometry and topology is often accomplished by writing the Cartesian tensors in their irreducible spherical forms using Clebsch-Gordon coefficients in order to identify the dipole, quadrupole, octupole, etc. components and has been extensively discussed in the literature \cite{jerph78.01,bance10.01}, as has their application to aromatic systems \cite{joffr92.01} and quantum graphs \cite{lytel13.01}.

\subsection{Solving quantum graphs with motifs.}\label{sec:motif}

The topological properties of a quantum graph are identical to those of the corresponding non-metric (relational) graph.  Such graphs contain a number of vertices, connected by directed or undirected edges.  Edges in quantum graphs have no direction.  Graphs contain vertices $V_{i}$ with $d_{i}$ edges connected to them (star vertices), where $d_{i}$ is the degree of vertex labeled $i$.  Graphs also contain loops and terminated edges.  A close examination of any graph shows that it may be constructed from a few fundamental graphs called \emph{motifs}.

\begin{figure}\centering
\includegraphics[width=1.7in]{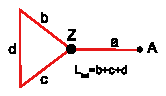}\includegraphics[width=1.7in]{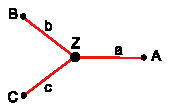}\\
\includegraphics[width=1.7in]{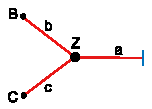}\includegraphics[width=1.7in]{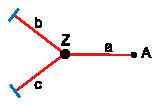}\\
\caption{The four primary motifs for constructing any graph.  The dark line at the end of an edge, such as those at the end of the edges labeled b and c in the lower right corner graph, indicates that the edge is terminated at an infinite potential.  Unterminated edges are labeled by an amplitude variable, indicating that the edge functions there are nonzero, as is the flux flowing into or out of the edge with that label.}
\label{fig:motifGraphs}
\end{figure}

The spectra of connected quantum graphs are the solutions to their secular equations, which always take the form of combinations of the secular functions of the motifs.  The motifs in Figure \ref{fig:motifGraphs} are the 3-star and lollipop graphs, and are sufficient to compute the states and spectra for all graphs.  For brevity, we limit graphs to those containing 3-stars, though the generalization is straightforward. An example is presented in \ref{sec:starAppendix}.

The nonlinearities of both the 3-star graph with edges terminated at infinite potential ($A=B=C=0$) and the lollipop with its stick terminated at infinite potential ($A=0$) have been calculated in the elementary QG model \cite{lytel13.01}. As isolated models of nonlinear, quantum confined systems, these are interesting structures because both topologies have intrinsic nonlinearities over half the fundamental limits.  The method of solution of arbitrary graphs comprised of the motifs is developed in detail in \ref{sec:solveWithMotifs}. With that machinery, all of the graphs discussed in the next section are computable. We move now directly to the results of the computations.  The interested reader may refer to the Appendices to learn how to compute graphs.

\section{Topological optimization of quantum graphs}\label{sec:topOpt}

\subsection{Topological and geometrical classes}

The character of quantum graphs comprised of star and lollipop motifs is dominated by the properties of the motifs.  Star and lollipop graphs have large intrinsic first and second hyperpolarizabilities, implying that composites containing stars and lollipops will have topological characteristics enabling geometric realizations with large hyperpolarizabilities.  Geometric constraints can reduce the dynamic range of the hyperpolarizability tensors by limiting the projections of the transition moments onto a specific external axis.  Further constraints, such as a closed topology with no external edges, can significantly alter the range of response for the graph \cite{shafe12.01}.

\begin{figure}\centering\scriptsize
\caption{Intrinsic nonlinearities of topological classes of quantum graphs. The first ($\beta_{xxx}$) and second ($\gamma_{xxxx}$) hyperpolarizabilities shown are the largest values for the geometries within the specific topological class.  The first hyperpolarizability tensor norm $\beta_{norm}$ is defined and calculated in the text, and is an invariant for the topological class. In all cases except closed loops, the maximum value of $\beta_{xxx}$ is equal to $\beta_{norm}$, indicating that the topology allows the graph to assume its best configuration for the xxx component, which usually means that the yyy component vanishes.  Loops are so tightly constrained that it is impossible for a loop to have one of its diagonal components at zero when the other is nonzero.}\label{fig:resultsTable}
\newcolumntype{S}{>{\centering\arraybackslash} m{1cm} } 
\begin{tabular}{S S S S S S }
  \hline\hline
Graph  & Geometry & Topology & $\beta_{norm}$ & $\left| \beta_{xxx} \right|$ & $\gamma_{xxxx}$ \\
  \hline\hline
\\
\includegraphics{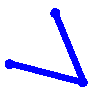} & bent wire & line & 0.172 & 0.172 & -0.126 to 0.007 \\
\hline
\\
\includegraphics{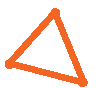} & triangle & loop & 0.086 & 0.049 & -0.138 to 0 \\
\hline
\\
\includegraphics{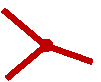} & 3-star & 3-fork & 0.58 & 0.58 & -0.138 to 0.3 \\
\includegraphics{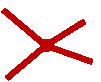} & 4-star & 4-fork & 0.53 & 0.53 & -0.125 to 0.27 \\
\includegraphics{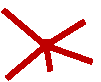} & 5-star & 5-fork & 0.51 & 0.51 & -0.11 to 0.26 \\
\includegraphics{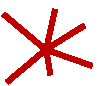} & 6-star & 6-fork & 0.51 & 0.51 & -0.11 to 0.26 \\
\includegraphics{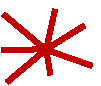} & 7-star & 7-fork & 0.51 & 0.51 & -0.11 to 0.26 \\
\hline
\\
\includegraphics{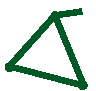} & lollipop & star-loop & 0.62 & 0.62 & -0.12 to 0.20 \\
\includegraphics{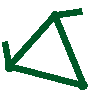} & bull & star-loop & 0.53 & 0.53 & -0.09 to 0.20 \\
\includegraphics{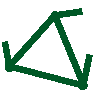} & lollipop bull & star-loop & 0.51 & 0.51 & -0.09 to 0.19 \\
\includegraphics{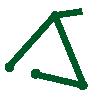} & lollipop & 3-fork & 0.33 & 0.33 & -0.11 to 0.13 \\
\includegraphics{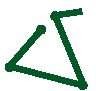} & lollipop & line & 0.17 & 0.17 & -0.09 to 0.006 \\
\hline
\\
\includegraphics{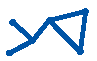} & barbell & 2-fork lollipop & 0.54 & 0.54 & -0.104 to 0.214 \\
\includegraphics{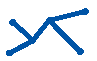} & barbell & dual 2-fork & 0.43 & 0.43 & -0.13 to 0.22 \\
\includegraphics{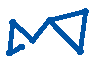} & barbell & star-loop & 0.41 & 0.41 & -0.07 to 0.11 \\
\includegraphics{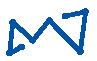} & barbell & line & 0.14 & 0.14 & -0.085 to 0.006 \\
\includegraphics{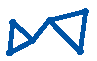} & barbell & loop & 0.11 & 0.11 & -0.1 to 0.002 \\
\hline\hline
\end{tabular}
\end{figure}

Wires, loops, and stars have spectra that are (more or less) evenly spaced.  Wires and loops have fixed eigenvalue spacing, whereas three-prong star graphs have fixed spacing between so-called root separators that divide the spectrum into cells of equal width, each containing a single eigenvalue.  The variation in spectra enabled by altering the lengths of the prongs of the star graph are due precisely to the variability between levels permitted by the root boundaries, but the achievement of any desired eigenvalue separation, such as that achieved in the Monte Carlo studies that generated near unity maxima, is not possible in a single star graph.  But as shown in this paper, many of the composite graphs in Figure \ref{fig:resultsTable} have nonlinearities larger than the three-prong star.  These same graphs have nonuniform root separators, and certain topological combinations of edge lengths enable variable level spacing that more closely resembles that achieved in the sum-rule-constrained Monte Carlo studies.  We anticipate that a sufficiently complex graph may be devised such that the level spacing of the most significantly-contributing levels could be near-optimum for achieving the maximum nonlinearity.

Figure \ref{fig:resultsTable} summarizes the results of the Monte Carlo study for all of the topologies discussed in this paper.  The Figure shows classes of graphs with the same topology but different geometry, the N-stars, and reveals that the nonlinearities are quite similar.  Figure \ref{fig:resultsTable} also shows a class of lollipop-like geometries.  The simple lollipop, the bullgraph, and the lollipop bull have similar nonlinearities, as their topologies are essentially identical.  But opening a corner of the lollipop turns the graph into a 3-star but with one bent prong. The star ensures the nonlinear is not small, but the geometry reduces the nonlinearities below those for the actual 3-star.  Opening the star turns the lollipop into a bent wire graph whose nonlinearity is now that of a wire, not a star.  Similar remarks hold for the class of barbell graphs:  Those with the star motif have about the same optical response, while opening both stars converts the graph to a wire and dramatically lowers the nonlinearity.  Finally, closing both stars into loops produces the greatest reduction in the response because, as with the loop graph, the ground state is now a zero energy state with constant flux throughout the graph.

\subsection{Topological clusters}

\begin{figure}\centering
\includegraphics[width=3in]{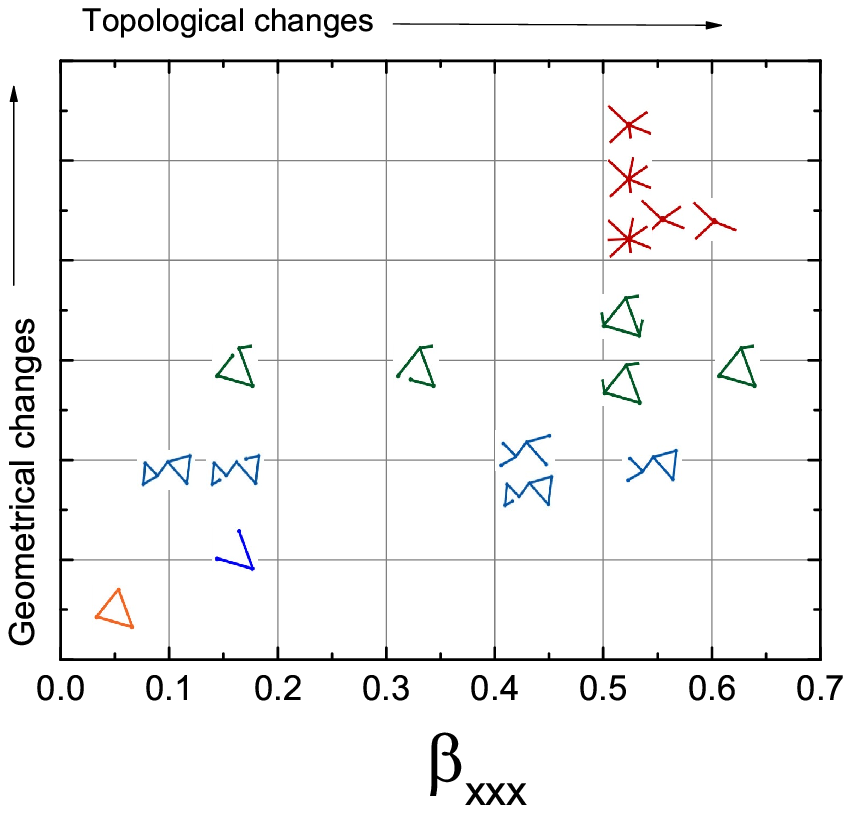}\\
\includegraphics[width=3in]{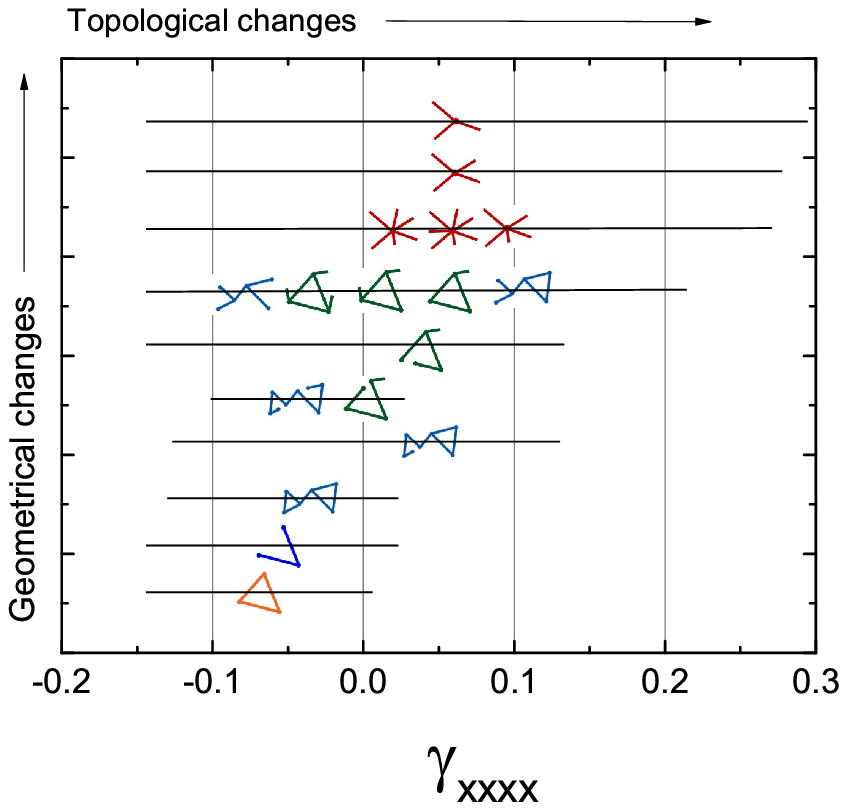}\\
\caption{Complete range of the first (top) and second (bottom) hyperpolarizability (horizonal axis) for the loop, wire, and star topologies (vertical axis) with various geometries for each. The vertical bins that change with $\beta_{xxx}$ (and $\gamma_{xxx}$) show that graphs with similar topologies have essentially the same hyperpolarizability.}
\label{fig:betaGammaRanges}
\end{figure}

Figure \ref{fig:betaGammaRanges} summarizes in pictorial form the relationship between the topology of the graph and its first and second hyperpolarizability for a collection of loops, wires, and stars.  The range for $\beta_{xxx}$ is always $-1\leq\beta_{xxx}\leq 1$, and is symmetric in absolute value around zero.  Graphs with positive $\beta_{xxx}$ may be rotated by $\pi$ to yield the identical negative value.  The top figure therefore displays only positive values.  Graphs placed along common vertical lines have the same topology but different geometry; graphs placed on the same horizontal line have the same geometry but different topology. A close examination reveals that loop and wire graphs without a star vertex cluster together with lower nonlinearities, while all graphs containing a star vertex cluster with much higher nonlinearities.  Yet all of these graphs have superscaling spectra.  For example, the spectrum of loops of length L is $E_{n}(loop)=2\pi^2 n^2/L^2$, while that of a wire of length L is $E_{n}(wire)=\pi^2 n^2/2L^2$.  The difference is that the eigenstates of the loop are doubly degenerate and have a zero energy ground state with constant flux (i.e., $n=0,\pm 1,\pm2\ldots$ for loops), while the eigenstates of wires are nondegenerate with a positive energy ground state.  The loop topology actually restricts the range of the transition moments, limiting the nonlinearities.  The wire topology has no such restriction.  But here, topology limits the response of the wire by creating cancellations in the state overlap of the transition moments, since eigenstates must oscillate continuously across the wire, creating both negative and positive overlap regions.

\subsection{Topological shifts in similar geometries}

The global properties of the hyperpolarizability tensors are thus determined by the topology of the graphs, while the local properties, such as the projections onto a fixed external axis, are determined by the geometry of a particular realization of the graph.  For a given Monte Carlo run on a specific topology, a complete sampling of possible geometries yields the ranges of the first and second hyperpolarizability, as well as the contributions to these tensors from their spherical components.  Graphs with identical shapes but different topologies necessarily have different spectra, though the projection of their edges onto a fixed external axis could be similar.  Topological shifts alter the spectra, changing both the values and the energy-level spacing; these factors set the limits on the maximum achievable hyperpolarizability in the graph, even when the geometry is optimized for that graph.

To study a particular typological class, we sample its configuration space using Monte Carlo methods. The coordinates of the edges that define the shape are selected at random; and, the intrinsic first and second hyperpolarizabilities calculated.  The distribution of results over many configurations provides insights into the relationship between a topological class and its nonlinear-optical properties.

\begin{figure}\centering
\includegraphics[width=3.1in]{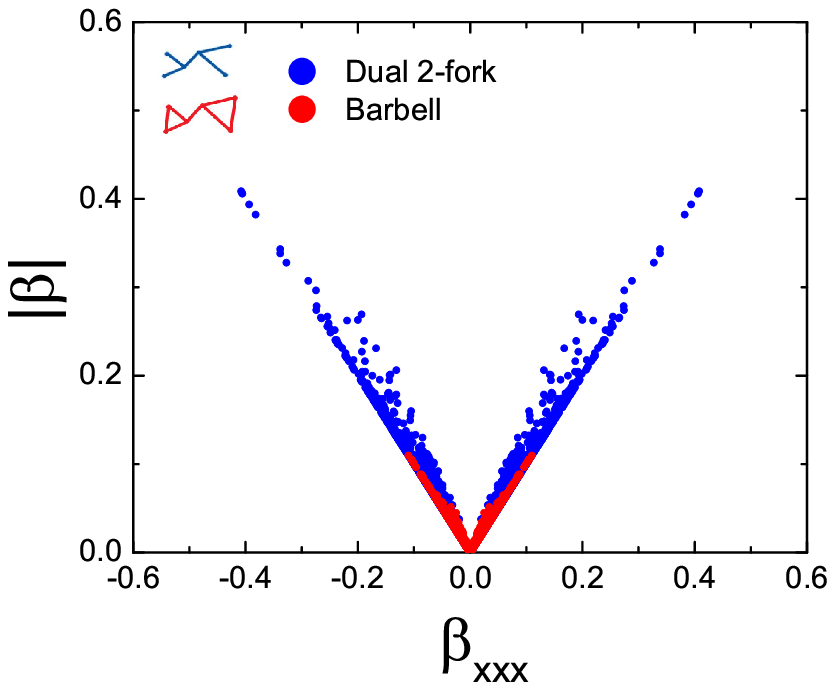}
\includegraphics[width=3.1in]{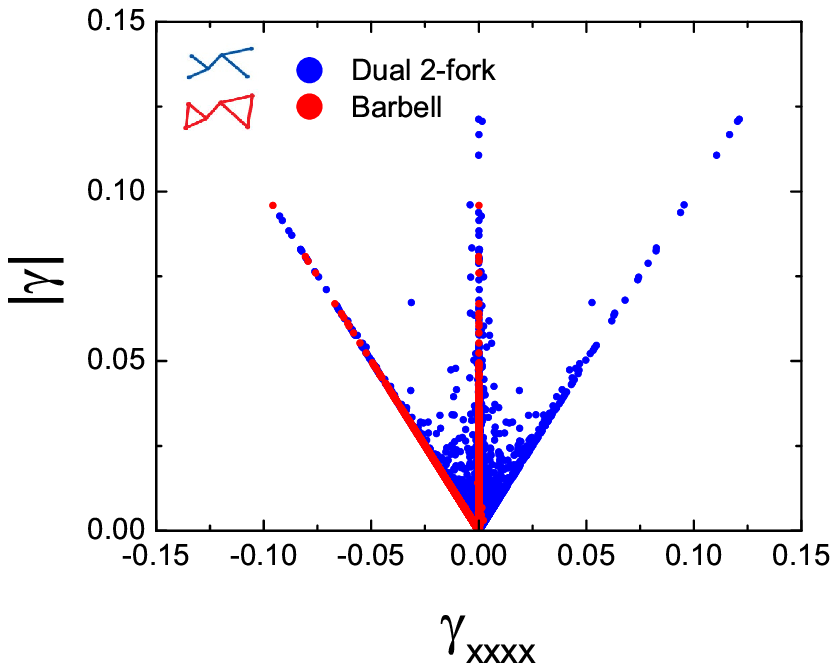}\\
\caption{Hyperpolarizability tensors and their norms for the barbell graph with two open ends (star-to-star) and two closed ends (bells) for a large sampling of shapes using a Monte Carlo method.  The profound change in the nonlinear response due to the topological change from a closed dual-loop configuration to a geometrically similar one that is isomorphic to two back-to-back 3-star graphs is self-evident.}
\label{fig:barbellTensors}
\end{figure}

Figure \ref{fig:barbellTensors} illustrates the approach when applied to two distinct topologies that span the same geometries:  Two realizations of a barbell, one containing closed bells and the other having two open bells.  The former contains two 3-star vertices but the stars are closed into loops, and the entire graph is sealed, as explained earlier in the paper.  The open barbell graph has two open stars connected in such a way that flux travels across the structure, rather than around the loops.  The energy levels of the graphs are quite different, and so are the hyperpolarizability tensors.

\begin{figure}\centering
\includegraphics[width=3in]{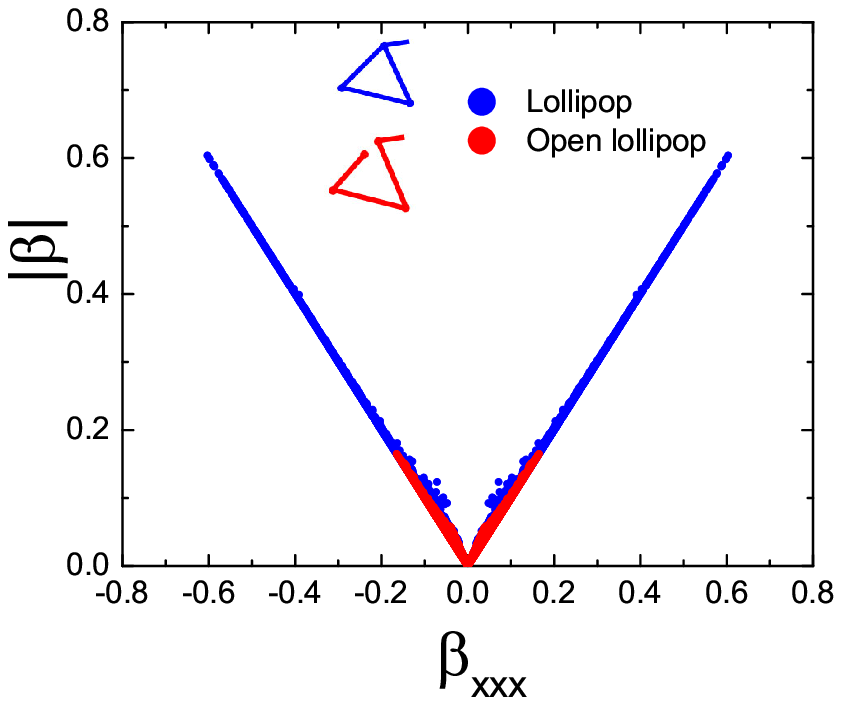}
\includegraphics[width=3in]{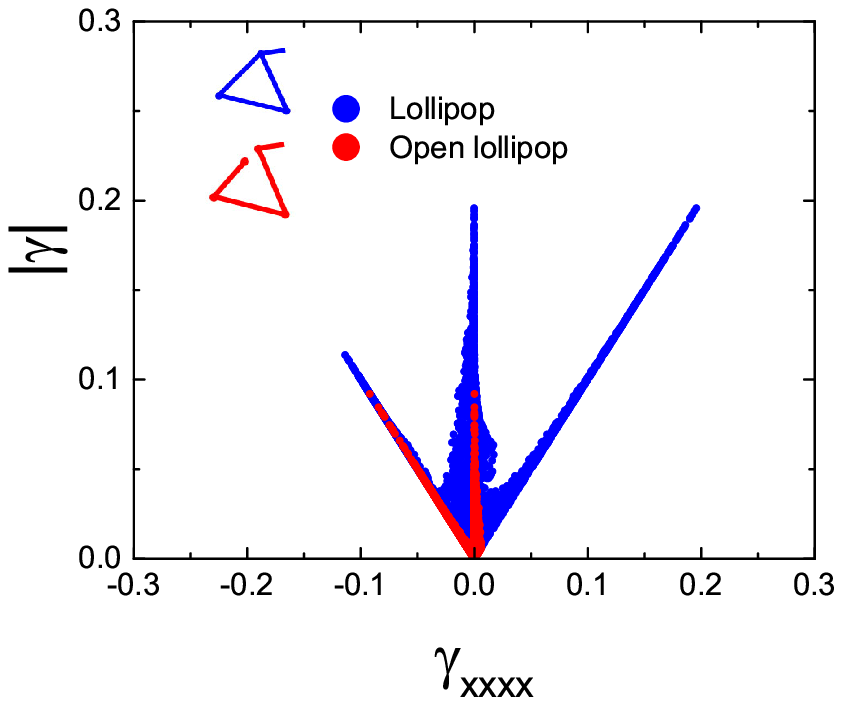}\\
\caption{Hyperpolarizability tensors and their norms for the lollipop graph and 4-wire lollipop-like graph for a large sampling of shapes using a Monte Carlo method.  The profound change in the nonlinear response due to the topological change from a star-based configuration to a geometrically similar one that is a bent wire graph is self-evident.}
\label{fig:lollipopTensors}
\end{figure}

Figure \ref{fig:lollipopTensors} applies the same approach to two distinct topologies that span the same lollipop geometry.  The lollipop graph contains a star vertex and a loop, and it has the largest first hyperpolarizability predicted for any of the star-containing graphs in this paper.  The  other a topological four-wire graph.  The energy levels of the graphs are quite different, and so are the hyperpolarizability tensors.

\subsection{Geometrical tuning}

\begin{figure}\centering
\includegraphics[width=4in]{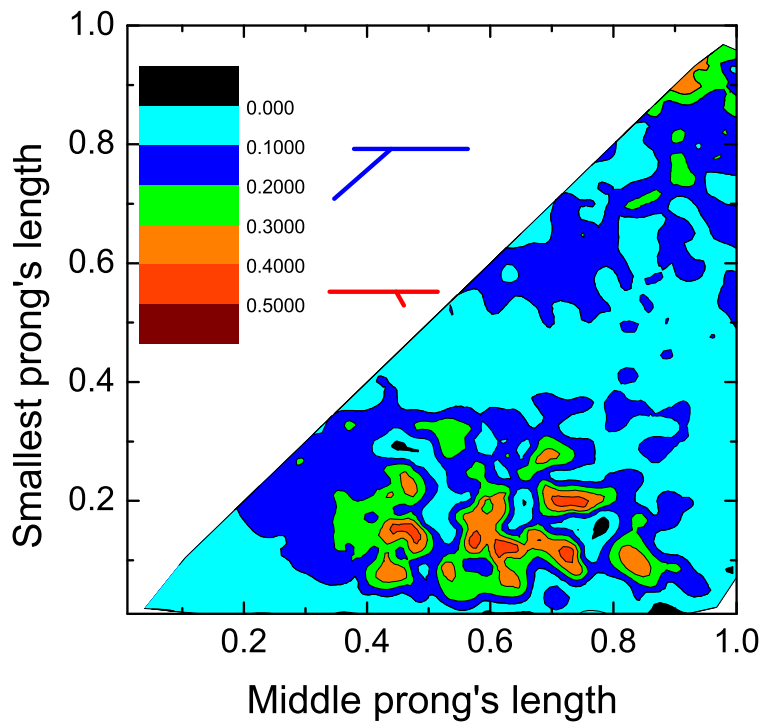}
\caption{Contour plot of the largest values of $\beta_{xxx}$ for 3-star graphs as a function of prong lengths.  The largest prong always has length unity (since the results are scale-independent).  The length of the middle prong ranges from zero to unity, while that of the shortest prong ranges from zero to a maximum equal to the middle prong. The angles each makes with the longest prong are random.  Each pair (short, middle) of prong lengths has a set of angles where $\beta_{xxx}$ is near zero, but only the optimum pairs (short, middle) can generate large $\beta_{xxx}$ for special sets of angles.  The inset shows the shape with the largest (red) and smallest (blue) $\beta_{xxx}$.}
\label{fig:starProngDist}
\end{figure}

\begin{figure}\centering
\includegraphics[width=4in]{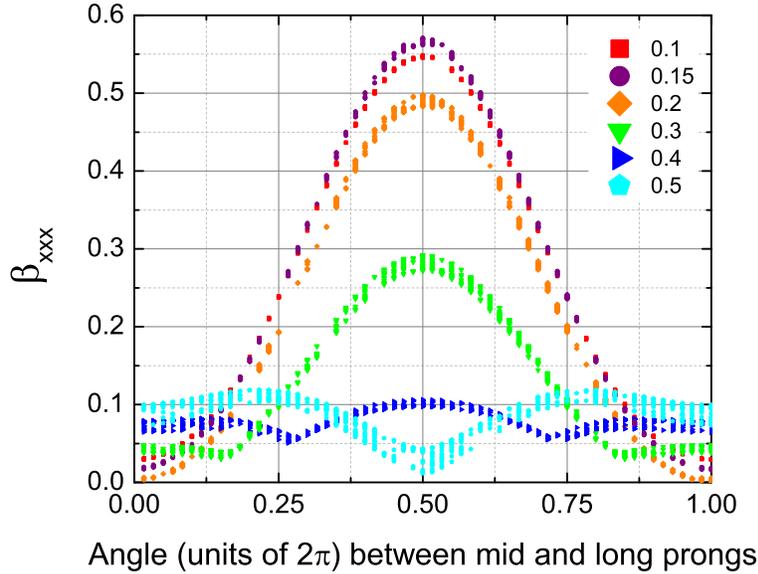}
\caption{Variation of $\beta_{xxx}$ with the angle between the middle prong of length 0.6 and the long prong of length unity for several short prong lengths.  The vertical range of points for a specific curve representing a short prong length is for the full range of angles of the small prong.  The key feature determining the strength of the nonlinearity is the antiparallel middle and large prongs, with a short prong at any angle.  A short prong permits the largest flux to move across the graph without diverting any of it into another direction.  Increasing the short prong length dramatically decreases the nonlinear response.  At a short prong length of 0.3 or greater, the nature of the angular dependence changes.}
\label{fig:thirdProngVariationBeta}
\end{figure}

Graphs containing star motifs have superscaling spectra, but their wavenumbers are no longer uniformly spaced.  Instead, the wavenumbers fall between evenly spaced \emph{root boundaries}\cite{pasto09.01}.  Energies have a quasi-quadratic spectrum, with level spacings that may be tuned by altering the geometrical properties of the graph.  Figure \ref{fig:starProngDist} illustrates the impact of edge length variations in 3-star graphs.  The relative edge lengths set the energy spectrum of the graph and also contribute to their projection onto an external x-axis once their angular positions are specified.  For a given set of prong lengths, the value of $\beta_{xxx}$ will vary over a range as the angles between the prongs change.  However, for each set of prong lengths, there will be one set of angles for which $\beta_{xxx}$ is maximum.  The figure was constructed so that the largest values were plotted on the top. For example, stars with prongs $(1,0.6,0.13)$ appear to have the largest $\beta_{xxx}$, but this is true only if the angles take on specific values.  Underneath the contours showing the greatest values for this set of prong lengths, there are points with lesser values corresponding to the same prong lengths but suboptimum angles; this is evident from Figure \ref{fig:thirdProngVariationBeta}.  The inset in Figure \ref{fig:starProngDist} shows the shapes with the largest values (red), as well as one with a much smaller value(blue).

The ideal star graph has its longest prong (of length one) and its second longest prong (of length $\sim 0.6$ antiparallel along $x$ and the shortest prong (of length $\sim 0.13$) at any angle.  Interestingly, a straight wire along $x$ would have zero $\beta_{xxx}$, while a bent wire could have $\beta_{xxx}\sim 0.172$ but no larger.  The attachment of a single short prong away from a center of symmetry converts the graph to a topology that generates one of the largest intrinsic values to date.

The star topology ensures that the flux may be distributed such that state overlap for the transition moments is maximized while flux is simultaneously conserved. The third prong in the graph mimics a discontinuity across the other two prongs due to a finite $\delta$ potential.  The addition of a prong to a wire to create a star is topologically equivalent to dressing a wire with a finite $\delta$ potential, a so-called \emph{dressed graph} that has recent been analyzed \cite{lytel13.04}.

We conclude that the effect on the hyperpolarizabilities of topological changes in quantum graphs is to induce spectral changes that favor higher response as well as allow geometrical tuning that can optimize the response through shape.  This latter point was carefully studied in Ref \cite{lytel13.01} for a class of loop graphs and a 3-star using a spherical tensor method \cite{jerph78.01,bance10.01,zyss94.01}.

\subsection{Approaching the fundamental limits}

\begin{figure}\centering
\includegraphics[width=3in]{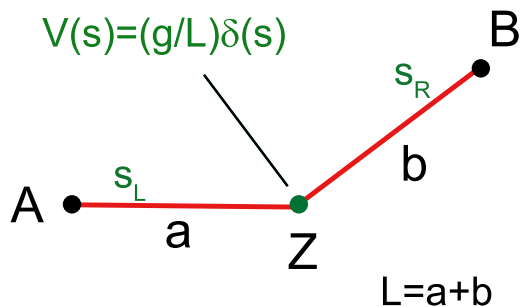}
\caption{A quasi-one dimensional quantum graph with a variable potential $(g/L)\delta(s)$ located between the endpoints.  This dressed graph has one of the largest nonlinearities (0.705) of any structure to date, nearly equal to the potential limit of $0.7089$.}\label{fig:deltaMotif}
\end{figure}

\begin{figure}\centering
\includegraphics[width=3.4in]{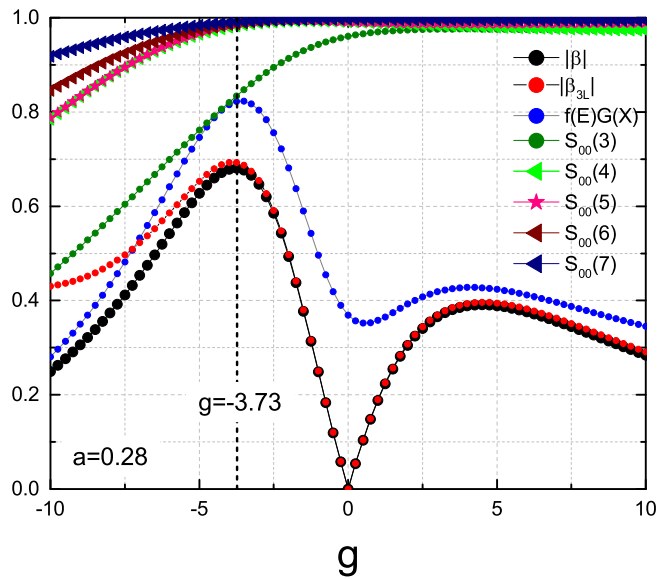}
\caption{Scaling of the full, three level, and extreme three level expressions for $\beta_{xxx}$ of a linear system with  a $\delta$ potential of strength g.  The three level model becomes exact at the maximum of $\beta_{xxx}$, which is the three level Ansatz in action.  The extreme three level model is nearly $20\%$ larger at the maximum, however.  Also shown are the values of the truncated sum rule $S_{00}$ with 3,4,5,6, and 7 terms.  The deviation of the models is clearly due to the requirement that the sum rule requires more than three terms to converge.}\label{fig:betabeta3fGsumrules}
\end{figure}

Superscaling spectra derived from a potential energy yield first hyperpolarizabilities with a maximum of $0.7089$ rather than unity \cite{zhou07.02,shafe13.01,ather12.01,lytel13.04}.  If only three states were required for the sum rule, the extreme three level model would be exact, and we would have $E=0.5$ but with the same (universal) value of $X\sim 0.79$. However, these models always require more than three states to satisfy the sum rules, as they must when the maximum is below unity.  Figure \ref{fig:deltaMotif} is a dressed quantum graph which nearly hits the maximum value of $\beta_{xxx}=0.7089$, while Fig. \ref{fig:betabeta3fGsumrules} illustrates the phenomena \cite{lytel13.04}.  For this example, E converges to a value of about 0.45.  The sum rules require a minimum of four states to converge, though the three level model is nearly exact at the maximum.  This is a general feature of nonlinear optical systems whose origin has been recently analyzed and discussed in detail \cite{kuzyk14.01}.

\section{Conclusions}\label{sec:outlook}

The ideal energy spectrum for maximizing the nonlinear optical response of any quantum structure scales as some positive power of the state number.  This has been previously established by Monte Carlo simulations valid for any non-relativistic many body Hamiltonian with random spectra and transition moments constrained by sum rules.  Heuristically, we argued that the theory of fundamental limits is also highly suggestive that so-called superscaling spectra are required if the optical nonlinearities are to bridge the large (factor of 30) gap between all presently known intrinsic molecular responses and the fundamental limit.

Quantum graphs--dynamical models of electrons confined to the edges of a mathematical graph on which a Hamiltonian operates--have exactly the superscaling spectra required to maximize the nonlinear optical response.  The distribution of energies is set by the graph topology, while the geometry may be altered to change the transition moments in an advantageous way.  Topology also affects the transition moments, in that it determines the flow of electrons in the graph and the shape of the eigenstates.  State shape and state overlap determine the signs and magnitudes of the transition moments. Consequently, graphs having distinct topological classes are expected to have similarly sized nonlinear optical responses.

In this paper, we presented a model of the generalized one electron quantum graph as a nonlinear optical structure, showed how to compute the hyperpolarizabilities, and invoked a general motif method to aid in calculating the spectra and states of more complex graphs containing star vertices, as well as loops and wires.  We provided a set of rules for calculating any graph, taking into account both degeneracies from rationally-related edges as well as the appearance of multiple sets of eigenstates arising from subgraphs.  We also related the global properties of closed graphs to the appearance of a zero energy, constant amplitude ground state.  The quantitative results of a large Monte Carlo study of these graphs were presented in Section \ref{sec:topOpt}, with the conclusion that graphs containing star motifs have the largest optical nonlinearities of any model structure known to date.

The general methods for solving for the hyperpolarizabilities of quantum graphs that we previously developed could then be used with the states and energies from the motif analysis to calculate the cartesian and spherical tensor components of the first ($\beta$) and second ($\gamma$) hyperpolarizability to understand the impact of topology across geometrically equivalent graphs on the nonlinear optical tensors.  In particular, graphs with identical topologies have comparable maximum nonlinearities, while graphs with identical geometries but different topologies have far different maximum nonlinearities.  This behavior has been previously observed for bent wires and loops \cite{shafe12.01} so it is not surprising that it holds for star graphs and their extensions.  But with the advent of the star motif for constructing the spectral equations for complex graphs, we now have a fundamental explanation for both the similar, topological responses and the differences when topologies are altered so that the underlying secular spectral functions of geometrically similar graphs no longer resemble one another.  Scaling according to the theory of fundamental limits also holds across different star geometries, so long as the star motif is active within the graph so that its global properties are dominated by the star topology.  Interestingly, the addition of a star vertex to a loop creates the lollipop graph which has one of the largest intrinsic first hyperpolarizabilities of all graphs, despite the fact that the loop by itself has a nonlinearity that is over ten times smaller.  The star vertex is key to the synthesis of molecular systems modeled by the elementary quantum graph, as it appears to guarantee that a geometrically-unconstrained star topology will have a large, intrinsic first and second hyperpolarizability.

The one-electron elemental graph model is a simple but effective way to explore a wide range of states and transition moments enabled by a structure's Hamiltonian and boundary conditions, from the bottom up, i.e., by solving the equations of motion to determine the maximum hyperpolarizabilities of a topological class of graphs for comparison with the abstract theory of fundamental limits based upon the use of the Thomas-Reiche-Kuhn sum rules in a sum over states expansion of $\beta$ and $\gamma$.  We have argued that multi-electron models will differ vastly in their construction and the richness of their physics, but that their spectra and transition moments should reflect the global properties of the one-electron model.  Consequently, we expect the results presented in this paper to be valid for actual quantum confined systems.

\section*{Acknowledgments}
SS and MGK thank the National Science Foundation (ECCS-1128076) for generously supporting this work.

\appendix
\setcounter{figure}{0}
\renewcommand{\thefigure}{A\arabic{figure}}
%

\section{Solution of quantum graphs using motifs}\label{sec:solveWithMotifs}
The results in the paper were computed using a new method for solving an arbitrary quantum graph for its states and spectra.

\subsection{Star graph motifs}

The conservation of flux in a star graph leads to the reduced secular function $f_{star}$ \cite{lytel13.01,pasto09.01}, where
\begin{equation}\label{reducedSecStar}
f_{star}(a_{i})=\sum_{i=1}^{E}\cot{k_{n}a_{i}}
\end{equation}
for an E-pronged star with edges $a_{i}$.

For the 3-star with edges $a,b,c$, multiply the reduced secular function by $\sin{k_{n}a} \, \sin{k_{n}b} \, \sin{k_{n}c}$, a factor that is nonzero for irrationally-related edges, and we get the secular function $F_{star}(a,b,c)$:
\begin{eqnarray}\label{3starSecularF}
F_{star}(a,b,c) &=& \frac{1}{4}\left[\cos{k_nL_1} + \cos{k_nL_2}\right.\nonumber \\
&+& \left.\cos{k_nL_3} - 3\cos{k_nL}\right],
\end{eqnarray}
where $L=a+b+c$, $L_1=|a+b-c|$, $L_2=|a-b+c|$, and $L_3=|a-b-c|$.
We call equation (\ref{3starSecularF}) the canonical form of the 3-star secular function and will use it extensively in what follows.  The combination lengths are equivalent to the edge lengths, and we freely move back and forth between them.  For example, a star graph with edges $d,e,f$ will have a secular function $F_{star}(d,e,f)$ which may be written in the form of the right hand side of equation (\ref{3starSecularF}) with the set $(a,b,c)$ replaced by $(d,e,f)$ in the definition of the combination lengths.

The nature of the solutions to the secular equation $F_{star}=0$ for irrational lengths have been discussed at length in Ref. \cite{pasto09.01}, where a periodic orbit expansion was derived for the eigenvalues.  They are nondegenerate and lie one to a cell between root boundaries at multiples of $\pi/L$.  For our purposes, a set of solutions for any finite number of wave functions is easily found by numerically intersecting the two parts of the secular equation.  In this way, a set of nondegenerate eigenvalues may be obtained for arbitrary (but irrational) prong lengths.  Solutions may be found in Ref. \cite{lytel13.01}. The degenerate case is solved later in this section.

\begin{figure}\centering
\includegraphics[width=4in]{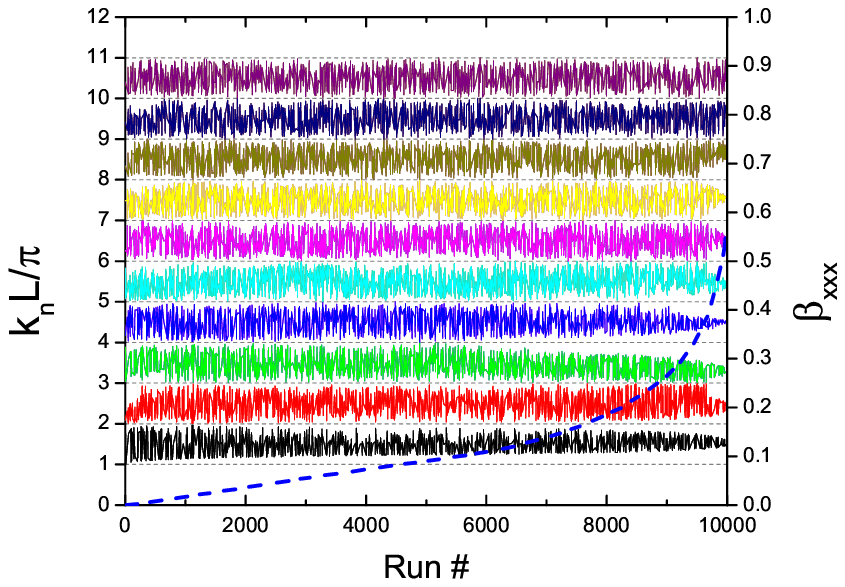}
\caption{Variation of the lowest ten eigenvalues of the 3-star graph for $10,000$ randomly generated samples, ordered by their $\beta_{xxx}$ values.  For the 3-star, the solutions to the secular equation lie between fixed root boundaries located at multiples of $\pi/L$, with L equal to the sum of all edges.  The three lowest eigenvectors asymptote to fixed values as $\beta_{xxx}$ (shown as a dashed curve) approaches its maximum value for the best geometry.}
\label{fig:starSpectrumBeta}
\end{figure}

In Figure \ref{fig:starSpectrumBeta}, the ten lowest eigenvalues are displayed for a Monte Carlo run where the ensemble members are ordered such that their maximum $\beta_{xxx}$ increases from left to right.  The numerical value of $\beta_{xxx}$ is displayed as the dashed curve and its value shown on the right axis in Figure \ref{fig:starSpectrumBeta}.  The eigenvalues each vary in a random way between their root separators as the geometry of the star is altered from left to right until the optimum geometries are attained.  When the maximum is approached, the lowest eigenvalues converge to well-defined values. The ratio $E\equiv E_{10}/E_{20}$ takes the value 0.4 for star graphs at their maximum.  Recall from Section \ref{sec:intro} that this three level model parameter is a measure of how \emph{close} the graph is to an optimum.

The final piece required to calculate hyperpolarizabilities are the transition moments.  These require the eigenstates for the graph, which are calculated from the individual edge states, per Eq. (\ref{xNM}). The edge states for the star graph have been detailed in Ref. \cite{lytel13.01}.

Consider now how to use the motifs to construct the secular functions for composite graphs containing stars.  For the star graph with three terminated ends, the secular function $F_{star}$ is exactly zero.  When the ends in the motifs are unterminated, the amplitudes at the ends are nonzero and there must be flux moving in or out of these ends, since flux is conserved in the graph.  This means that the secular function is no longer zero but is related to the flux entering or leaving the unterminated vertices.  For the fully unterminated 3-star motif in Figure \ref{fig:motifGraphs}, the canonical form of the edge functions is
\begin{eqnarray}\label{3starEdges}
\phi_n(s_{a}) &=& \frac{Z_n \sin k_n\left(a-s_{a}\right)+ A_{n}\sin k_{n}s_{a}}{\sin k_{n}a} \nonumber \\
\phi_n(s_{b}) &=& \frac{Z_n \sin k_n\left(b-s_{b}\right)+ B_{n}\sin k_{n}s_{b}}{\sin k_{n}b} \nonumber \\
\phi_n(s_{c}) &=& \frac{Z_n \sin k_n\left(c-s_{c}\right)+ C_{n}\sin k_{n}s_{c}}{\sin k_{n}c} \nonumber \\
&&
\end{eqnarray}
where each distance s on an edge is measured from the central vertex.  For unterminated ends, conservation of flux at the central vertex $Z$ produces the following secular equation relating the amplitudes at the ends and the central amplitude:
\begin{eqnarray}\label{3starSecUnterm}
Z_{n}F_{star}(a,b,c) &=& A_n\sin k_{n}b\sin k_{n}c \nonumber \\
&+& B_n\sin k_{n}a\sin k_{n}c \\
&+& C_n\sin k_{n}a\sin k_{n}b \nonumber
\end{eqnarray}
The right-hand side is the net flux through its unterminated vertices required to conserve flux at the central vertex.  If the ends are terminated, the right-hand side vanishes, reproducing the secular equation for a terminated star graph, $F_{star}(a,b,c)=0$.  For unterminated ends, equation (\ref{3starSecUnterm}) relates the amplitudes at the ends and at the central vertex through a single equation.

\begin{figure}\centering
\includegraphics[width=2.5in]{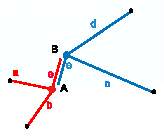}\includegraphics[width=2.5in]{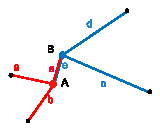}\\
\caption{The hybrid star-to-star graph formed from the union of two star motifs.}
\label{fig:star2star}
\end{figure}

We may use the 3-star motif to compute the secular equation of a graph consisting of two 3-star motifs.  Consider the graph in Figure \ref{fig:star2star} with two star vertices connected by a common prong. There are two central vertices connected by an edge, and each is a 3-star motif with two ends at zero amplitude.  The coupled amplitude equations are easy to write down using equation (\ref{3starSecUnterm}) for each star and appropriate relabeling the vertex amplitudes and edges to match those of the composite graph. They are
\begin{eqnarray}\label{star2starAmps}
A_{n}F_{star}(a,b,e)=B_{n}\sin k_na\sin k_nb\nonumber \\
B_{n}F_{star}(c,d,e)=A_{n}\sin k_nc\sin k_nd
\end{eqnarray}
The secular function for this graph is thus
\begin{eqnarray}\label{secularStar2star}
F_{star-star} &=& F_{star}(a,b,e)F_{star}(c,d,e) \\
&-& \sin{k_na}\sin{k_nb}\sin{k_nc}\sin{k_nd}\nonumber
\end{eqnarray}

The secular function in equation (\ref{secularStar2star}) may be rewritten in the following form:
\begin{eqnarray}\label{SecularFcnS2S2}
F_{sec} &=& -4\sin{k_{n}a}\sin{k_{n}b}\sin{k_{n}c}\sin{k_{n}d}\sin{k_{n}e}\nonumber \\
&-& 2\sin{k_{n}(a+b+c+d+e)} +\sin{k_{n}(a+b-c-d+e)}\nonumber \\
&-& \ \sin{k_{n}(a+b-c-d-e)}\nonumber +.5\sin{k_{n}(a+b+c-d+e)}\nonumber \\
&+& .5\sin{k_{n}(a+b+c-d-e)}+ .5\sin{k_{n}(a+b-c+d+e)} \\
&+& .5\sin{k_{n}(a+b-c+d-e)}+ .5\sin{k_{n}(a-b+c+d+e)}\nonumber \\
&+& .5\sin{k_{n}(a-b+c+d-e)}-.5\sin{k_{n}(a-b-c-d+e)}\nonumber \\
&-& .5\sin{k_{n}(a-b-c-d-e)}\nonumber
\end{eqnarray}

Using motifs, we have obtained the secular function for this back-to-back star graph in a few steps.  The amplitudes at the two vertices are easily calculated from Eq. (\ref{star2starAmps}).  The transition moments and hyperpolarizabilities are computed using the machinery from Section \ref{sec:QGreview}.  The generalization to graphs comprised of many stars is straightforward.

\subsection{Lollipop motifs}

For the lollipop graph, the secular function $F_{pop}(a,L_{tot})$ is \cite{lytel13.01}
\begin{eqnarray}\label{lollipopSecular}
F_{pop}(a,L_{tot}) &=& \frac{1}{2}\left[3\cos k_n \left(a+\frac{L_{tot}}{2}\right)\right.\nonumber \\
&-& \left.\cos k_n \left(a-\frac{L_{tot}}{2}\right)\right].
\end{eqnarray}
where $L_{tot}=b+c+d$ is the length of the loop and $a$ is the prong length.

The wavefunctions of the lollipop graph are a composite of two sets of wavefunctions, one set that is nonzero at the central vertex and on all edges, and one for wavefunctions that vanish at the origin and are exactly zero on the prong edge.  The first set correspond to the symmetric wavefunctions of a 3-sided bent wire (open at the central vertex) coupled to a nonzero prong wavefunction, while the second set correspond to the asymmetric wavefunctions of a 3-sided bent wire (open at the central vertex) with a zero prong wavefunction.  When another graph is attached to the prong, the loop-only wave functions go away and we're left with the wave functions satisfying the secular equation above \cite{lytel13.01}.

\begin{figure}\centering
\includegraphics[width=4in]{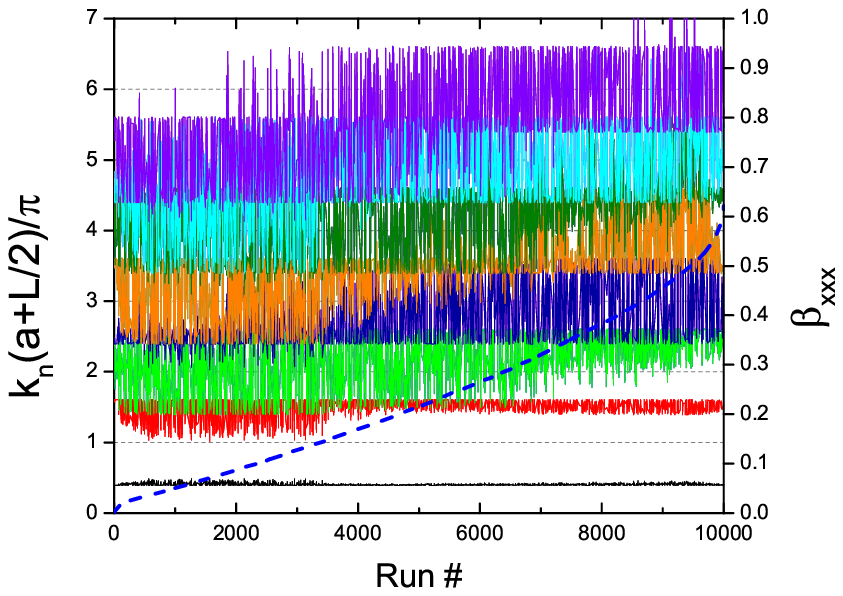}
\caption{Variation of the lowest eight momentum eigenvalues of the lollipop graph for $10,000$ randomly generated samples, ordered by their $\beta_{xxx}$ values. $\beta_{xxx}$ is shown as the dashed curve.}
\label{fig:lollipopSpectrumAndBeta}
\end{figure}

The spectrum of the star motif had uniform root boundaries between which all eigenvalues were found for any geometry of the graph.  This observation is not true for the lollipop graph, whose spectrum is a complex interleaving of two sets of disparate spectra as discussed above.  The spectrum for a Monte Carlo run of lollipops with variable geometry is illustrated in Figure \ref{fig:lollipopSpectrumAndBeta}, ordered by increasing $\beta_{xxx}$.  There are always well-defined boundaries between a given set of eigenstates for a fixed run, and that somewhere between runs  3000 and 4000, $\beta_{xxx}$ begins to climb, and the states jump to a different-looking pattern where the variation of the three lowest eigenvalues decreases rapidly and then converge to fixed values at the maximum hyperpolarizability, with a universal value of the energy ratio $E\sim 0.4$.

Since the maximum value of $\beta_{xxx}$ for lollipops is larger than that of the basic star graph, we might expect that further changes in the complexity of the spectrum of a graph could lead to even larger responses.  In complex graphs, the root boundaries may acquire an almost random structure to them, suggesting they might be {\em tunable} to provide the kinds of level spacing required to achieve maximum nonlinear responses.

Consider now the unterminated lollipop in Figure \ref{fig:motifGraphs}.  The exact expression for the flux in/out of the lollipop motif is
\begin{equation}\label{lollipopSecUnterm}
Z_{n}F_{pop}(a,L_{tot}) = A_{n}\cos k_{n}L_{tot}/2.
\end{equation}
The left-hand side is the total flux exiting the central vertex Z and entering the vertex A.  When $A=0$, the flux conservation equation becomes $F_{pop}(a,L_{tot})=0$.  This determines the eigenvalues of the terminated lollipop graph where there is flux moving on all of its edges but never exiting at vertex A.  As noted above, the terminated lollipop has an additional spectrum comprised of wave functions where there is exactly zero flux on edge a at all times, i.e., flux just circulates around the loop.  This set must be included in the total spectrum of the lollipop.

\begin{figure}\centering
\includegraphics[width=2.5in]{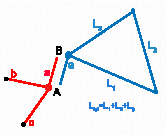}\includegraphics[width=2.5in]{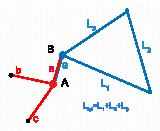}\\
\caption{The hybrid star-lollipop graph formed from the union of the star and lollipop motifs.}
\label{fig:lollipop2starstick}
\end{figure}

We are now in a position to use the secular functions for the star and lollipop motifs to solve for the secular equation of the combined graph in Figure \ref{fig:lollipop2starstick}.  We see from the Figure that we should connect the lollipop to the star such that vertex Z of the star in Fig \ref{fig:motifGraphs} is vertex A of the lollipop in the same figure. Relabeling the vertices and edges to match those of the composite graph, we get
\begin{eqnarray}\label{lollipop2starAmps}
A_{n}F_{star}(a,b,c) &=& B_{n}\sin k_{n}b\sin k_{n}c \\
B_{n}F_{pop}(a,L_{tot}) &=& A_{n}\cos k_{n}L_{tot}/2 \nonumber
\end{eqnarray}
Cross-multiplying (or setting the determinant of the coefficients to zero) yields the secular function $F_{pop-star}(a,b,c,d,e)$ for the eigenvalues of the star-stick lollipop graph:
\begin{eqnarray}\label{seclollipop2star}
F_{pop-star}(a,b,c,d,e) &=& F_{star}(a,b,c)F_{pop}(a,L_{tot}) \\
&-& \sin k_{n}b\sin k_{n}c\cos k_{n}L_{tot}/2 \nonumber
\end{eqnarray}
The solutions to $F_{pop-star}(a,b,c,d,e)=0$ are the eigenvalues of the star-stick lollipop graph.  The amplitudes $A_{n}$ and $B_{n}$ are then found from Eq. (\ref{lollipop2starAmps}). With these in hand, the hyperpolarizabilities for this class of graphs are calculated as described previously in Section \ref{sec:QGreview}.

\section{Computational rules for general graphs}\label{sec:compRules}
Every graph may be solved using the fundamental star and lollipop motifs in the same way as done for the previous composite graphs.  We wish to note some general rules for using motifs to solve graphs.  We will again limit the discussion to internal vertices of degree equal to three or less, but the generalization to internal vertices of arbitrary degree is straightforward and requires use of the N-star motifs for degree $N$.

The general method to writing down a secular equation for a graph with $V_{I}$ internal vertices is as follows:  (1) Label the vertices with their amplitudes $A,B,C,\ldots$ where flux flowing into or out of each is conserved and must flow along edges connected to the vertex and to other parts of the graph, (2) determine how the vertex and its edges overlays other vertices and their connected edges in order to identify the motifs comprising the graph, (3) use the secular functions in equation (\ref{3starSecUnterm}) and equation (\ref{lollipopSecUnterm}) with appropriately relabeled amplitudes to write a set of simultaneous equations relating the secular functions at an internal vertex to the connecting amplitudes via the motif equations, and (4) set the determinant of the amplitude matrix to zero to obtain the secular equation for the entire graph.  The secular equation of a graph is generally transcendental but is easily solved using numerical methods.

The secular equation provides a set of eigenvalues for the states of the graph where the amplitude of the particle on each edge is nonzero.  Graphs containing closed loops, such as the lollipop or barbell, can also have wavefunctions where the amplitude along a connecting or terminal edge is exactly zero, as was described for the lollipop motif in this section.  When multiple sets of wavefunctions are present, they must be ordered in energy and their eigenstates interleaved so that a complete set results for the graph.  Finally, graphs with no external connections, such as a barbell or triangle, will necessarily have a zero-energy eigenstate where the wavefunction over the entire graph is constant.  This ground state must be included in the spectrum in order to maintain completeness of the eigenstates.  For most composite graphs, there will not be any additional sets of spectra other than those from the secular equation.  Again, the rational case is an exception, allowing wavefunctions that vanish at the shared vertices and form exact half-periods over each edge.  These are straightforward to handle, should they arise, and do not require solution to any transcendental secular equation.

We realize that the rules presented here are not nearly as expeditious as are Feynman rules for particle interactions.  But they are general and enable any graph to be analyzed rapidly.  It will be true, however, that as the size of the graph grows, it will be challenging to solve the resultant secular equation, as there are likely to be many configurations with degenerate states.  Degeneracies require a bit of care to deal with, and are discussed next.  Moreover, it will be necessary to track the metric distance of each edge from a common origin in order to compute the transition moments.  For graphs with a dozen edges, this isn't too challenging, but one can imagine a graph having twenty edges in a rather irregular configuration being quite difficult to solve.  Fortunately, we are seeking quantum graphs with large nonlinearities that also reflect the physics or chemistry of simple molecular systems, and we have already discovered that the basic star and lollipop graphs have large nonlinearities.

\setcounter{figure}{0}
\renewcommand{\thefigure}{D\arabic{figure}}
\section{Handling degeneracies}\label{sec:degeneracies}

\begin{figure}\centering
\includegraphics[width=3.4in]{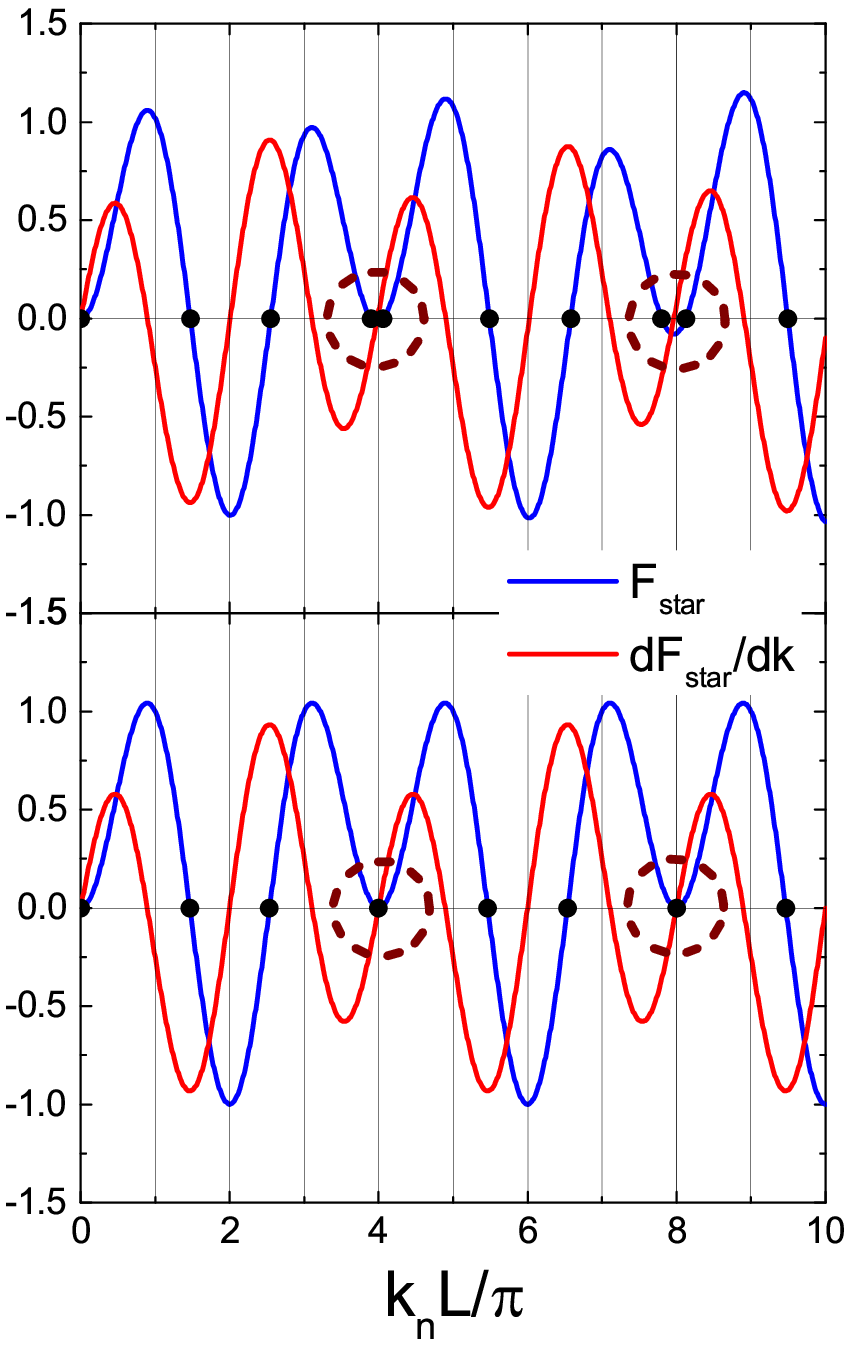}
\caption{Solutions to the secular equation of a 3-star graph with prong lengths that are irrationally-related (upper panel) but approach a rational relationship (lower panel).  The two roots enclosed by the dotted circles in the irrational case (one each on either side of a root separator) coalesce into a pair of degenerate roots (on a root separator) enclosed by the dotted circle, as the lengths become rationally-related  The transition is smooth in that there is no abrupt change in the nonlinear response of the graph as the edges become rationally-related. Units on the y-axis are arbitrary.}\label{fig:secular_nondeg_to_deg}
\end{figure}

Throughout this paper, the edges of the stars have been constrained to be irrationally-related.  This ensures the canonical form of the edge functions in equation (\ref{edgeFunctionAB}) may be used without reservation, as the edge functions never vanish at the central vertices.  We briefly examine rationally-related edges to show how to solve these, too.
The isolated star graph can have doubly-degenerate states for wavenumbers satisfying $k_{n}=n \pi/L$, with $L=a+b+c$ for certain values of the edges. If we write the edge functions for the three prongs as
\begin{eqnarray}\label{edgeStates3starDeg}
\phi_{n}^{(1)}(s_{1}) &=& A_{n}\sin{k_{n}(a-s_{1})}\nonumber \\
\phi_{n}^{(2)}(s_{2}) &=& B_{n}\sin{k_{n}(b-s_{2})} \\
\phi_{n}^{(3)}(s_{3}) &=& C_{n}\sin{k_{n}(c-s_{3})}\nonumber
\end{eqnarray}
then the amplitudes at the center satisfy
\begin{equation}\label{3starAmps}
A_{n}\sin{k_{n}a} = B_{n}\sin{k_{n}b} = C_{n}\sin{k_{n}c},
\end{equation}
and conservation of flux yields the secular function for an isolated star graph as
\begin{equation}\label{secDeg}
F_{star} = A_{n}\cos{k_{n}a} + B_{n}\cos{k_{n}b} + C_{n}\cos{k_{n}c}.
\end{equation}
If none of the sine functions in equation (\ref{3starAmps}) vanishes, then the wavenumbers satisfy the usual secular equation, (\ref{3starSecularF}).  The derivative of the secular function is
\begin{equation}\label{secDerivDeg}
-dF_{star}/dk = aA_{n}\sin{k_{n}a} + bB_{n}\sin{k_{n}b} + cC_{n}\sin{k_{n}c}.
\end{equation}
For irrationally-related edges, $F_{star}(k_{n})=0$ determines the nondegenerate eigenvalues $k_{n}$, and the derivative $dF_{star}/dk$ is never zero for $k=k_{n}$.  But when the edges are rationally-related, both the secular equation and its derivative will occasionally vanish for the same $k$, the doubly-degenerate eigenvalues.  When this occurs, Eq. (\ref{secDeg}) and Eq. (\ref{secDerivDeg}) may be used to extract amplitudes for a pair of orthogonal, degenerate states corresponding to the same eigenvalue, because the same secular equation holds for the degenerate case, as well \cite{pasto09.01}, as can be shown through a scattering matrix solution or simply by noting that the transition from irrationally-related edges to rationally-related ones is equivalent to an infinitesimal change in the arguments of Eq. (\ref{secDeg}).  Consequently, one may move from one case to the other by performing all the divisions used to derive Eq. (\ref{3starSecularF}) and then taking the rational limit.
To see how this comes about, examine the spectrum in Figure \ref{fig:secular_nondeg_to_deg}.  The vertical lines at $k_{n}=n\pi/L$ (with $L=a+b+c$) are the root separators, defining cells in which only one root may be found.  For certain values of the edges, there are roots on either side of a root separator that converge toward each other and meet at a separator (becoming degenerate roots) as the edge values are tweaked toward specific ratios.  When this happens, all three terms in Eq. (\ref{3starAmps}) vanish, and they also vanish in the derivative of the secular equation (which is why the roots are doubly-degenerate).  A single degenerate root has a pair of eigenstates whose amplitudes are determined by the secular Eq. (\ref{secDeg}) and the requirement that the pair of degenerate states are orthogonal.  If the edge coefficients are labeled $(A_{1}B_{1}C_{1})$ and $(A_{2}B_{2}C_{2})$, the orthogonality condition is $aA_{1}A_{2}+bB_{1}B_{2}+cC_{1}C_{2}=0$.  A suitable set of coefficients may then be determined from this and the secular relations, with $A_1=1$, $A_2=1$, $C_1=1$ as the roots converge to $k_{r}$ as follows:
\begin{eqnarray}\label{degAmpCoeff}
B_{1} &=& -\frac{\cos{k_{r}c}+\cos{k_{r}a}}{\cos{k_{r}b}}\nonumber \\
C_{2} &=& -\frac{a\cos{k_{r}b}-bB_{1}\cos{k_{r}a}}{c\cos{k_{r}b}-bB_{1}\cos{k_{r}c}}\nonumber \\
B_{2} &=& -\frac{\cos{k_{r}a}+C_{2}\cos{k_{r}c}}{\cos{k_{r}b}}
\end{eqnarray}
where the cosines will take the values $\pm 1$ as their arguments each approach their own multiple of $\pi$.
Even when there are no degeneracies for a given set of rationally-related edges, it is possible that one or two of the sine functions in equation (\ref{3starAmps}) could vanish.  (If all three vanish, then the root is a degenerate root boundary).  In this case, the amplitudes may still be obtained by using the amplitude equation (\ref{3starAmps}) and the secular equation.  For example, suppose a given solution to the secular equation $k_{m}$ satisfies $\sin{k_{m}a}=0$, but that $\sin{k_{m}b}\neq 0$ and $\sin{k_{m}c}\neq 0$.  Then (\ref{3starAmps}) and the secular equation yield the singlet solution set
\begin{eqnarray}\label{oneSineZero}
B_{m} &=& C_{m}\frac{\sin{k_{m}c}}{\sin{k_{m}b}} \\
A_{m} &=& -B_{m}\frac{\sin{k_{m}(b+c)}}{\cos{k_{m}a}\sin{k_{m}c}}\nonumber \\
\end{eqnarray}
The single unknown coefficient $C_{m}$ is determined by normalization, of course.  When two sine functions vanish, say $\sin{k_{m}a}=0$ and $\sin{k_{m}b}=0$, the solutions are even easier to obtain.  Then (\ref{3starAmps}) yields $C_{m}=0$ and $A_{m}=-B_{m}\cos{k_{m}b}/\cos{k_{m}a}$.  This solves the degenerate case for any relationship among the edges.  The extension of this analysis to graphs comprised of star motifs is straightforward, but it provides no additional information to that obtained from the nondegenerate case.

\section{Scaling to $N\geq 4$ star vertices}\label{sec:starAppendix}

\begin{figure}\centering
\includegraphics[width=1.3in]{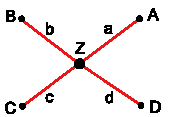}\includegraphics[width=2.1in]{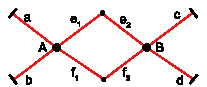}\\
\caption{Left:  4-star motif with four open edges carrying flux to/from another motif to which their edges might be attached.  Right: a bubble graph comprised of two 4-star motifs.}
\label{fig:4starMotif}
\end{figure}

Referring to Figure \ref{fig:4starMotif}, we easily generalize equation (\ref{3starEdges}) to arrive at the 4-edge equivalent of equation (\ref{3starSecUnterm}) and get
\begin{eqnarray}\label{4starSecUnterm}
Z_{n}F_{4star}(a,b,c,d) &=& A_n\sin{k_{n}b}\sin{k_{n}c}\sin{k_{n}d} \nonumber \\
&+& B_n\sin{k_{n}a}\sin{k_{n}c}\sin{k_{n}d} \\
&+& C_n\sin{k_{n}a}\sin{k_{n}b}\sin{k_{n}d}\nonumber \\
&+& D_n\sin{k_{n}a}\sin{k_{n}b}\sin{k_{n}c}\nonumber
\end{eqnarray}
where the secular function for the 4-star graph is given by the four-edge version of equation (\ref{reducedSecStar}), rationalized by the denominators.  After some simple algebra, the secular function may be written as
\begin{eqnarray}\label{4starSecularF}
F_{4star}(a,b,c,d) &=& \frac{1}{2}\left[\sin{k_{n}(a+b)}\cos{k_{n}(c-d)}\right. \nonumber \\
&+& \left. \cos{k_{n}(a-b)}\sin{k_{n}(c+d)}\right.\nonumber \\
&-& \left. \sin{k_{n}(a+b+c+d}\right].
\end{eqnarray}
This form may be used as a motif to solve composite graphs with several four-edge vertices, such as the bubble graph shown in Figure \ref{fig:4starMotif}, for which the secular function may again be written down by inspection:
\begin{eqnarray}\label{4starBubbleSecularF}
&& F_{4bubble}(a,b,c,d,e_1,e_2,f_1,f_2)= \\
&& F_{4star}(a,b,L_1,L_2)F_{4star}(c,d,L_1,L_2)\nonumber \\
&-& \sin{k_{n}a}\sin{k_{n}b}\sin{k_{n}c}\sin{k_{n}d}(\sin{k_{n}L_{1}}+\sin{k_{n}L_{2}})^{2}\nonumber
\end{eqnarray}
where $L_{1}=e_{1}+e_{2}$, $L_{2}=f_{1}+f_{2}$ are the two (sequential) bubble edges.
Generalizing to any number of edges is straightforward.  We note here that adding prongs to a 3-star simply changes its geometry, not its topology.  The global behavior of the nonlinearities is therefore similar to that of the 3-star.

\end{document}